\documentclass[preprint,floats,aps,showpacs,epsf]{revtex4}
\usepackage{amssymb}

\usepackage{epsfig}

\usepackage{amsmath}
\usepackage{psfrag}
\usepackage{color}
\usepackage{graphicx}
\usepackage{subfigure}

\newcommand{\be}{\begin{equation}}
\newcommand{\ee}{\end{equation}}
\newcommand{\bea}{\begin{eqnarray}}
\newcommand{\eea}{\end{eqnarray}}

\newcommand{\ri}{\mbox{i}}

\def\chiud{\chi^{\rm ud}_{11}}
\def\chiuo{\chi^{\rm ud}_{12}}
\def\chidd{\chi^{\rm d}_{11}}
\def\chido{\chi^{\rm d}_{12}}

\begin{document}
\title{Modelling Magnetic Fluctuations in the Stripe Ordered State}

\author{R. M. Konik$^1$, F. H. L. Essler$^2$,  and A.  M. Tsvelik$^1$}
\affiliation{$^1$ Department of  Condensed Matter Physics and Materials Science,
Brookhaven National Laboratory, Upton, NY 11973-5000, USA\\
$^2$ The Rudolph Peierls Centre for Theoretical Physics,
Oxford University, 1 Keble Road, Oxford  OX1 3NP, UK} 
\date{\today}

\begin{abstract}
The nature of the interplay between 
superconductivity and magnetism in
the cuprates remains one of the fundamental unsolved problems in high
temperature superconductivity. Whether and how these two phenomena are
interdependent is perhaps 
most sharply seen in the stripe phases of various
copper-oxide materials.  These phases, involving a mixture of spin and
charge density waves, do not yet admit a complete, overarching
theoretical treatment.  However aspects of this problem can be
analyzed. In this work, we focus on the magnetic side of stripe 
physics.  To this end, we study a simple model of 
a stripe-ordered phase consisting of an array of alternating
coupled doped and undoped two-leg Hubbard-like ladders.  To obtain the
magnetic response, we employ already available dynamical susceptibilities of
the individual two-leg ladders and treat the interladder coupling
in a random phase approximation.  Strikingly, we find two possible
scenarios for the ordered state induced by the coupling between
ladders: the spin modulation can occur either {\it along} or {\it
perpendicular} to the direction of the stripes.  These two
scenarios are differentiated according to different microscopic
realizations of the component doped ladders.  
However inelastic neutron scattering experiments on the two stripe
ordered cuprates, La$_{1.875}$Ba$_{0.125}$CuO$_4$ 
and $\rm La_{2-x}Sr_xCuO_4$, do not readily
distinguish between these two scenarios due to
manner in which stripes form in these materials.

\end{abstract}

\maketitle

\sloppy
\section {Introduction}
$\rm La_{2-x}Ba_{x}CuO_4$ is the material where high temperature
superconductivity was first discovered by Bednorz and M\"uller in
1986 \cite{bedmull}.  At $x=0.125$ this copper oxide sees an anomalous
suppression  of $T_c$ \cite{moodenbaugh} 
which has been argued to be coincident with
static stripe order, a unidirectional static charge and spin density
wave.  Support for the existence of this order has been found both in
neutron scattering \cite{fujita,tran_nat} and x-ray \cite{abba}
measurements. Beyond $\rm La_{1.875}Ba_{0.125}CuO_4$, the most prominent
cuprate exhibiting static stripe order is neodymium-doped LSCO, 
$\rm La_{1.6-x}Nd_{0.4}Sr_xCuO_4$ \cite{nickelates}.
However evidence for ``dynamic stripes'', a phenomenon characterized
by short-range CDW order
and incommensurate low (but finite) energy magnetic excitations, is
found in a number of materials. 
Such excitations have been observed both in $\rm YBa_2Cu_3O_{6+x}$
crystals \cite{dai,mook,keimer1,hinkov} and in
$\rm La_{2-x}Sr_xCuO_4$ \cite{shirane,shirane1} over a range of dopings. 

Magnetic order appears in two different guises in these
copper-oxides. In neutron measurements on untwinned crystals of
$\rm YBa_2Cu_3O_{6.6}$ exactly {\it two} incommensurate 
low energy peaks are seen \cite{mook,keimer1}.  While initial observations 
of the phonon anomaly suggested that the
peaks were located {\it perpendicular}
to the direction of the stripes \cite{mook1}, later measurements of the same anomaly \cite{pint}
suggested the opposite conclusion, that the magnetic order was found {\it parallel} to the stripes.
The origin of magnetic order in $\rm La_{1.875}Ba_{0.125}CuO_4$ and $\rm La_{2-x}Sr_{x}CuO_4$
is similarly ambiguous but for different reasons.  In these materials,
{\it four} peaks in the neutron scattering intensity are observed.
This doubling in the number of peaks corresponds to a doubling of the
unit cell in the $\rm La$-based materials.  Each cell spans two
copper-oxide planes where the stripes in each plane are orientated at
$90^o$ relative to one another.  The doubling obscures the orientation of the magnetic
relative to the charge order, again opening 
up the the possibility that magnetic order may conceivably arise
not {\it perpendicular} but {\it parallel} to the stripes.

Previous theoretical efforts aimed at deriving the magnetic excitation
spectrum in the stripe ordered state have treated the doped regions as
structureless magnetic voids
\cite{vojta,uhrig,carlson}. Such an approach ignores the internal
dynamics of these regions. In this paper we attempt to take these
dynamics into account.  We do so by adopting a simplified model of
static stripes suggested for $x=1/8$ doped LBCO by
Tranquada et al. \cite{tran_nat}, where the unit cell in a single plane
contains one undoped and one doped two-leg ladder. 

The presence of the doped two-leg ladders (in lieu of magnetically
inert voids) has two important consequences.  First and foremost it
allows us to develop two scenarios for magnetic ordering.  In the
first scenario, magnetic order develops perpendicular 
to the stripe direction (or in our model, perpendicular to the ladder).  
In the second scenario, magnetic order develops parallel to the
stripe/ladder, a scenario, as we have indicated,
that cannot be excluded necessarily from either $YBa_2Cu_3O_{6.6}$ due to
ambiguities in measurements of the phonon anomaly nor from the $\rm La$-based compounds because
of their bi-plane structure of stripe ordering.  In our model of
coupled ladders, one scenario is favoured over the other on the basis
of particular non-universal features in the spin response of an
individual doped ladder \cite{note}.  This non-universality then implies that at
least in the context of our model, one scenario is not fundamentally
more natural than the other. 

The second consequence of note that flows from our model is a natural
explanation for the $\pi$ phase shift concomitant with the
incommensurate magnetic order.  In models where the doped striped
regions are ignored, the undoped parts of the copper-oxide plane are
connected via effective {\it ferromagnetic} couplings.  Such couplings
must be employed if the correct incommensurate order is to be
produced. Instead here, we show that a model of {\it anti-ferromagnetically} 
connected doped and undoped ladders is able
to produce the $\pi$-phase shift.  In effect, we show how to generate
dynamically the ferromagnetic coupling between undoped regions. 

A fundamental assumption underlying the model we are
analyzing is that stripes are not merely a low energy phenomenon
but rather exist over a large range of energies. Support for this
view may be derived 
from inelastic neutron scattering experiments \cite{tran_nat,tran_rev,xu},
where a strong inelastic signal between $50{\rm meV}$ and 
$100{\rm meV}$ has been attributed to arise from stripe correlations. 
However is not yet fully understood \cite{kiv} how to
reconcile such a stripe based picture with the existence of nodal
quasiparticles established in angle resolved photoemission data
  \cite{ARPES}.  And while stripe correlations may exist
at higher energies, they are certainly not an isolated phenomena.  Typically,
higher energy inelastic neutron scattering observations \cite{tran_nat,xu} only see broad features,
indicating at the least, strong damping.  

While our model pertains primarily to magnetic order at
1/8 doping where the incommensurate ordering wavevector equals $Q_s =
\pi(1 \pm 1/4,1)$ or $Q_s = \pi(1,1 \pm 1/4)$, it is also capable of
describing other values of incommensuration. In the second scenario of
ordering presented below, 
the incommensuration results from the position of the low lying
quasi-coherent mode on the doped ladder which is itself a 
linear function of doping. While this requires the assumption that 
that doping adds holes to the stripe without changing the distance
between stripes, it is perhaps a useful step towards a
description of the striped phase in $La_{2-x}Sr_{x}CuO_4$ for $0.055 <
x < 0.125$ where parallel stripe order appears with incommensuration
linear in $x$. 

While we do not address directly the origin of superconductivity, 
a particularly attractive feature of this model is that
superconductivity arises naturally from the strong pairing correlations
present in doped ladders \cite{Fabrizio,Fisher,AFK,konricetsv}.   In order to judge the
applicability of such a model it is first important to analyze its implications
for the magnetic dynamics of the striped phase.  This is the aim of the
present paper.

\section{The Model}
\begin{figure*}
\begin{center}
\noindent
\epsfysize=0.35\textwidth
\epsfbox{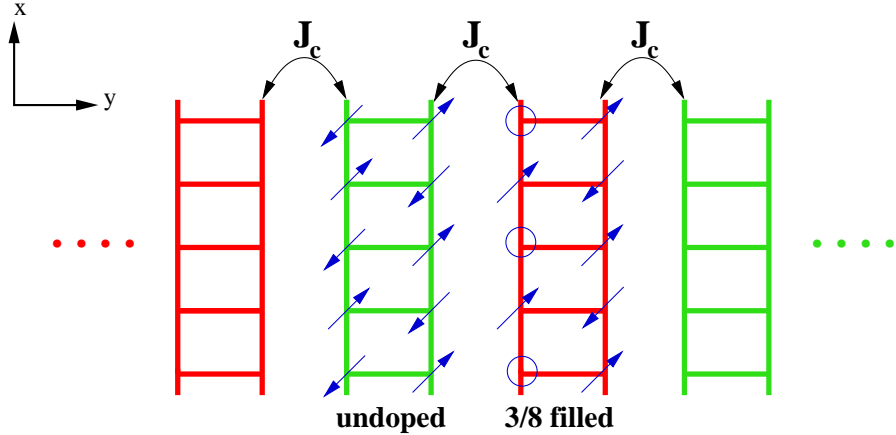}
\end{center}
\caption{A schematic of an alternating infinite array of coupled 
half-filled and doped ladders.  We take the coupling $J_c$ to
be antiferromagnetic.}
\label{fig:ladders}
\end{figure*}

\begin{figure}
[ht]
\begin{center}
\epsfxsize=0.7\textwidth
\epsfbox{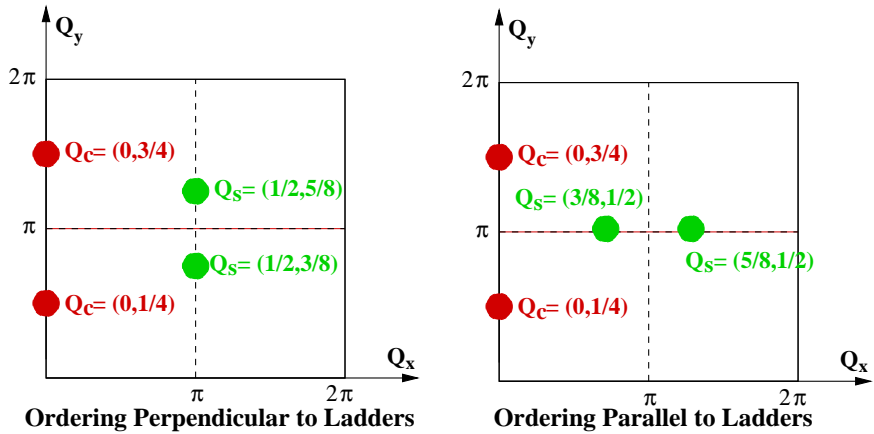}
\end{center}
\caption{Two possible scenarios for magnetic and charge long range order in the
coupled ladder model.  The wavevectors are marked in units of $2\pi$.}
\label{fig:orders}
\end{figure}

The basic model underlying our calculations is illustrated in
Fig. \ref{fig:ladders}: we have an array of alternating doped and
undoped Hubbard-like ladders. The charge gap in the undoped ladders is
taken to be very large. As a result the dominant interaction between
ladders is antiferromagnetic superexchange $J_c$.
As is clear from the above discussion, the experimentally
observed charge order with commensurate wave vector
\begin{equation}
{\bf Q}_{\rm c}=(0,\pm\frac{\pi}{2})
\end{equation}
is built into our model from the very beginning. The issue we want to
address is the static spin order that develops upon coupling the
ladders together as well as the spin dynamics. It is widely believed
that magnetic long-range order develops at
\begin{equation}
{\bf Q}_{\rm s}=(\pi,\pi\pm\frac{\pi}{4}),
\end{equation}
that is, perpendicular to the stripes.
On the basis of the analysis presented below,
we suggest that an alternative scenario is possible. Here magnetic
long-range order develops {\sl along} the direction of the stripes at
wave vectors
\begin{equation}
{\bf Q}_{\rm s}=(\pi\pm\frac{\pi}{4},\pi).
\end{equation}
The two scenarios are illustrated in Fig. \ref{fig:orders}. 

\section{Analysis of the Magnetic Response}
The basic ingredients of our approach are dynamical susceptibilities of
the two types of ladders. As our subsequent analysis is based on a
random phase approximation (RPA) in the interladder couplings, this is
the only information required. It turns out that the results obtained
in such an approach display a certain robustness with respect to
changing the microscopic details of the model. This allows us to identify
prominent features of the magnetic response which we believe to be
insensitive of the particular approximations we employ.

The dominant interladder coupling is taken to be antiferromagnetic
superexchange of strength $J_c$ (see Figure \ref{fig:ladders}), which
is induced by virtual hopping processes between doped and undoped
ladders. The matrix susceptibilities for the undoped (U) and doped (D)
ladders are expressed in terms of the matrices
\begin{eqnarray}\label{eIIIi}
M_U(\omega,q_x,q_y) &=& 
\left(
\begin{array}{cc}
\chi^{\rm ud}_{11}(\omega,q_x) & 
e^{iq_y a}\chi^{\rm ud}_{12}(\omega,q_x) \\
e^{-iq_y a}\chi^{\rm ud}_{21}(\omega,q_x)
 & \chi^{\rm ud}_{22}(\omega,q_x) 
\end{array}\right)\cr\cr
M_D(\omega , q_x, q_y) &=& 
\left(
\begin{array}{cc}
\chi^{\rm d}_{11}(\omega,q_x) & 
e^{iq_y a}\chi^{\rm d}_{12}(\omega,q_x) \\
e^{-iq_y a}\chi^{\rm d}_{21}(\omega,q_x) 
& \chi^{\rm d}_{22}(\omega,q_x).
\end{array}\right)
\end{eqnarray}
via
\begin{figure}
[ht]
\begin{center}
\epsfxsize=0.45\textwidth
\epsfbox{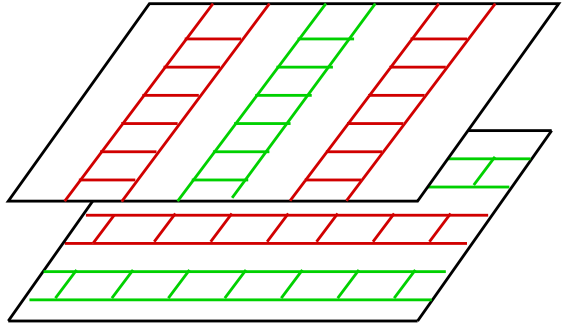}
\end{center}
\caption{Stacking of planes of ladders.}
\label{fig:planes}
\end{figure}
\begin{eqnarray}\label{eIIIii}
\chi_a(q_x,q_y,\omega ) &=& {\rm Tr} M_a\cdot K\ ,\quad a=U,D\ ,
\end{eqnarray}
where $K$ is defined by 
\begin{equation}\label{eIIIiii}
K = \left(
\begin{array}{cc}
1 & 1 \\
1 & 1
\end{array}
\right).
\end{equation}
Here $\chi_{11}=\chi_{22}$ marks correlations along the legs of the ladder while
$\chi_{12}=\chi_{21}$ describes correlations of the ladder rungs.

The coupling between the ladders is then taken into account in RPA. In
the matrix notation introduced above this amounts to 
\begin{eqnarray}\label{eIIIiv}
\chi^{\rm RPA}_{\rm 2D}(\omega,q_x,q_y) &=& 
{\rm Tr}\Bigl((1+M_DJ)M_U(1-JM_DJM_U)^{-1}K\Bigr)\ ,\nonumber\\
&&+{\rm Tr}\Bigl((1+M_UJ)M_D(1-JM_UJM_D)^{-1}K\Bigr)\ ,
\end{eqnarray}
where $J$ is a matrix given by
\begin{equation}\label{eIIIv}
J = \left(
\begin{array}{cc}
0 & e^{-iq_ya}J_c \\
e^{iq_ya}J_c & 0
\end{array}
\right).
\end{equation}
The scattering function for the coupled ladders is then
$S(q_x,q_y,\omega ) \sim -{\rm Im} \chi^{RPA}_{2D}(\omega,q_x,q_y)$.

Long range magnetic ordering occurs when $\chi^{2D}(\omega
=0,q_x,q_y)$ develops a singularity at some $Q_x$ and $Q_y$.  
In the RPA the development of the singularity is equivalent to the
vanishing of ${\rm Det} (1-JM_UJM_D)$, which gives
\begin{eqnarray}\label{eIIIvi}
0&=& (1-J_c^2\chidd\chiud)^2 
+ J_c^4\bigg((\chido\chiuo)^2-(\chido\chiud)^2 
- (\chidd\chiuo)^2\bigg) \nonumber\\
&& -2J_c^2\chido\chiuo\cos(4 q_y).
\end{eqnarray}
In the above, both $\chi^{\rm d}_{ij}$ and $\chi^{\rm ud}_{ij}$ are
functions of only $q_x$ and $\omega$, while  $q_y$ only appears in the final
cosine.  Once we have the doped and 
undoped ladder susceptibilities in hand, we will
readily be able to determine the value of the transverse wavevector, $q_y$, at which
order arises.  We will find two scenarios, one with order at $q_y = \pi \pm \pi/4$,
and one with order at $q_y = \pi$.  One of our main conclusions is that
which scenario is realized depends on
the details of the ladder susceptibilities.

\section{Ladder Susceptibilities: General Structure}

\subsection{Susceptibility of the undoped ladders}
The low energy spectral weight of the undoped ladders is
concentrated around $q_x=q_y=\pi$ and the susceptibility
displays a modulation along the y-direction by the factor
$(1-\cos(q_y))$ \cite{lake}. As long as we restrict our attention to energies
below the two magnon continuum (which dominates the response at
$q_y=0$), we can express the susceptibilities of the undoped ladders
in the form
\bea
\chi^{\rm ud}_{ab}(\omega,q_x)&=&
\chi^{\rm ud}(\omega,q_x)\left(
\begin{array}{cc}  
1 & -1\\
-1 & 1
\end{array}
\right)_{ab}\ ,
\eea
where
\begin{equation}
\chi^{\rm ud}(\omega,q_x)=\frac{Z(q_x)}{\omega^2-\epsilon^2(q_x)}.
\end{equation}
The magnon dispersion relation, $\epsilon (q_x)$, is taken from Ref. 
\onlinecite{barnes}, 
\begin{equation}
\epsilon (q_x) = J((1.89\cos(q_x/2))^2+(.507\sin (q_x/2))^2
 + (1.382\sin (q_x))^2)^{1/2}.
\end{equation}
The residue $Z(q_x)$ can be inferred from Ref. \onlinecite{SU05}. We
use the following simple, approximate fit,
$Z(q_x) = 3J(0.65\sin^2(q_x/2) + 0.27)$.

\subsection{Susceptibility of the doped ladders}
\begin{figure}
[ht]
\begin{center}
\epsfxsize=0.45\textwidth
\epsfbox{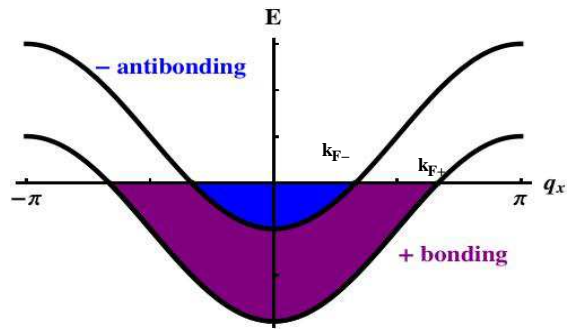}
\end{center}
\caption{Band structure of Hubbard-like ladders. $\pm$ denote
  bonding and antibonding bands, respectively, and the chemical
  potential generically leads to partial filling of both bands.}
\label{fig:bandladder}
\end{figure}

In order to infer the susceptibilities of the doped ladders it is
useful to recall the band structure. There are two bands corresponding
to bonding (+) and antibonding (-) fermions respectively,
$c_{\pm,\sigma}=(c_{1,\sigma}\pm c_{2,\sigma})/\sqrt{2}$. Generically
both bands will cross the chemical potential, leading to four Fermi
wave numbers $-k_{F\pm}$ and $k_{F\pm}$ as is illustrated in
Fig.\ref{fig:bandladder}.
The mapping to the bonding and antibonding picture implies the
following decomposition of the doped susceptibilities $\chi_{ab}^{\rm
  d}$ 
\bea
\chi^{\rm d}_{ab}(\omega,q_x) &=& 
 \chi^{\rm d}_{\rm intra}(\omega,q_x)+
\chi^{\rm d}_{\rm inter}(\omega,q_x)\left(
\begin{array}{cc}  
1 & -1\\
-1 & 1
\end{array}
\right)_{ab}.
\eea
Here $ \chi^{\rm d}_{\rm intra}(\omega,q_x)$  and
$\chi^{\rm d}_{\rm inter}(\omega,q_x)$ denote the parts of the
susceptibility involving only fermions within the same band and
fermions of both bands respectively.

The band structure further dictates that low-energy spin excitations
occur at $q_x\approx 0, \pm 2k_{F+}, \pm 2k_{F-}$ in 
$\chi_{\rm   intra}^{\rm d}$ and at $q_x\approx \pm k_{F+} \mp k_{F-},
\pm k_{F+} \pm k_{F-}$ in $\chi_{\rm   inter}^{\rm d}$
respectively. At low energies, we therefore can write
\begin{eqnarray}
\chi^{\rm d}_{\rm intra}(\omega,q_x)&=&
\chi^{\rm d}_0(\omega,q_x)+ \chi^{\rm d}_{2k_{F+}}(\omega,q_x)+ 
\chi^{\rm d}_{2k_{F-}}(\omega,q_x)\ ,\\
\chi^{\rm d}_{\rm inter}(\omega,q_x)&=&
\chi^{\rm d}_{k_{F+}+k_{F-}}(\omega,q_x)+ 
\chi^{\rm d}_{k_{F+}-k_{F-}}(\omega,q_x).
\end{eqnarray}
Here the single magnon weight (though only quasi-coherent due to
the gapless charge excitations of the doped ladder) is found in
$\chi^d_{k_{F+}+k_{F-}}$. The remaining contributions represent two excitation
continua.

\subsection{Magnetic Instability}
In terms of the inter and intraband susceptibilities of the doped
ladders the RPA instability condition (\ref{eIIIvi}) reads
\bea
1 = 4J_c^2\chi^{\rm ud}(\omega,
q_x)\left[\sin^2(2q_y)\chi_{\rm intra}^{\rm d}(\omega,q_x) +
  \cos^2(2q_y)\chi^{\rm d}_{\rm inter}(\omega,q_x)\right].
\label{RPA} 
\eea
This form makes it obvious that there are three possible sources for a
magnetic instability: it can be driven by 1) the interband
susceptibility of the doped ladders, $\chi^{\rm d}_{\rm inter}(\omega,q_x)$; 2) 
the intraband susceptibility of the doped ladders, $\chi^{\rm d}_{\rm intra}(\omega,q_x)$;
or finally, 3) by the susceptibility of the undoped ladder, $\chi^{\rm ud}(\omega, q_x)$. In
the first case the ordering occurs at $q_y=\pi\pm\frac{\pi}{4}$; in the 
second case 
at $q_y~{\rm mod}~\pi=0,\frac{\pi}{2}$; and in the final case at $q_y = \pi$. Let us further
elaborate on these three possibilities.

\subsubsection{Scenario I: Instability due to $\chi^{\rm d}_{\rm intra}(\omega,q_x)$}

In this scenario, the ordering occurs at 
\begin{equation}
{\bf Q}^I_{s}=(\pi,\pi \pm \frac{\pi}{4}).
\end{equation}
Here the ordering arises from predominance of the two-particle scattering continuum
over the single particle magnon.
In this scenario, the specific form of the doped susceptibilities, based upon treating the doped
ladders as a manifestation of the $SO(6)$ Gross-Neveu model, are given in Appendix A.

\subsubsection{Scenario II: Instability due to $\chi^{\rm d}_{\rm inter}(\omega,q_x)$}

The ordering occurs at
\begin{equation}
{\bf Q}^{II}_{s}=(\pm(k_{F+}+k_{F-}),\pi).
\end{equation}
Here the ordering arises from the spectral weight associated with
the single particle magnon on the doped ladder, found near wavevector
$D=k_{F+}+k_{F-}$.  (For 1/8-doped LBCO, $D=\frac{3\pi}{4}$.
In this scenario this spectral weight overwhelms that
of the two-particle continuum in the doped ladders (as encoded in $\chi^{\rm d}_{\rm intra}(\omega,q_x)$).
To then treat this case, we imagine that the susceptibility of the doped ladder comes
solely from the single particle magnon as discussed in further detail in Appendix A.

\subsubsection{Scenario III: Instability due to $\chi^{\rm ud}(\omega,q_x)$}

In this final scenario, the ordering occurs at the commensurate wavevector
\begin{equation}
{\bf Q}^{III}_{s}=(\pi,\pi).
\end{equation}
Here the ordering arises because the spectral weight of the undoped ladder dominates.
It is however the least relevant scenario for describing neutron scattering experiments
on the cuprates and so will not be explored in detail here.

\renewcommand{\thesubfigure}{}
\renewcommand{\subfigcapskip}{-4mm}
\begin{figure*}
\begin{center}
\subfigure[$\omega=6meV$]{
\epsfysize=.45\textwidth
\includegraphics[height=2.9cm]{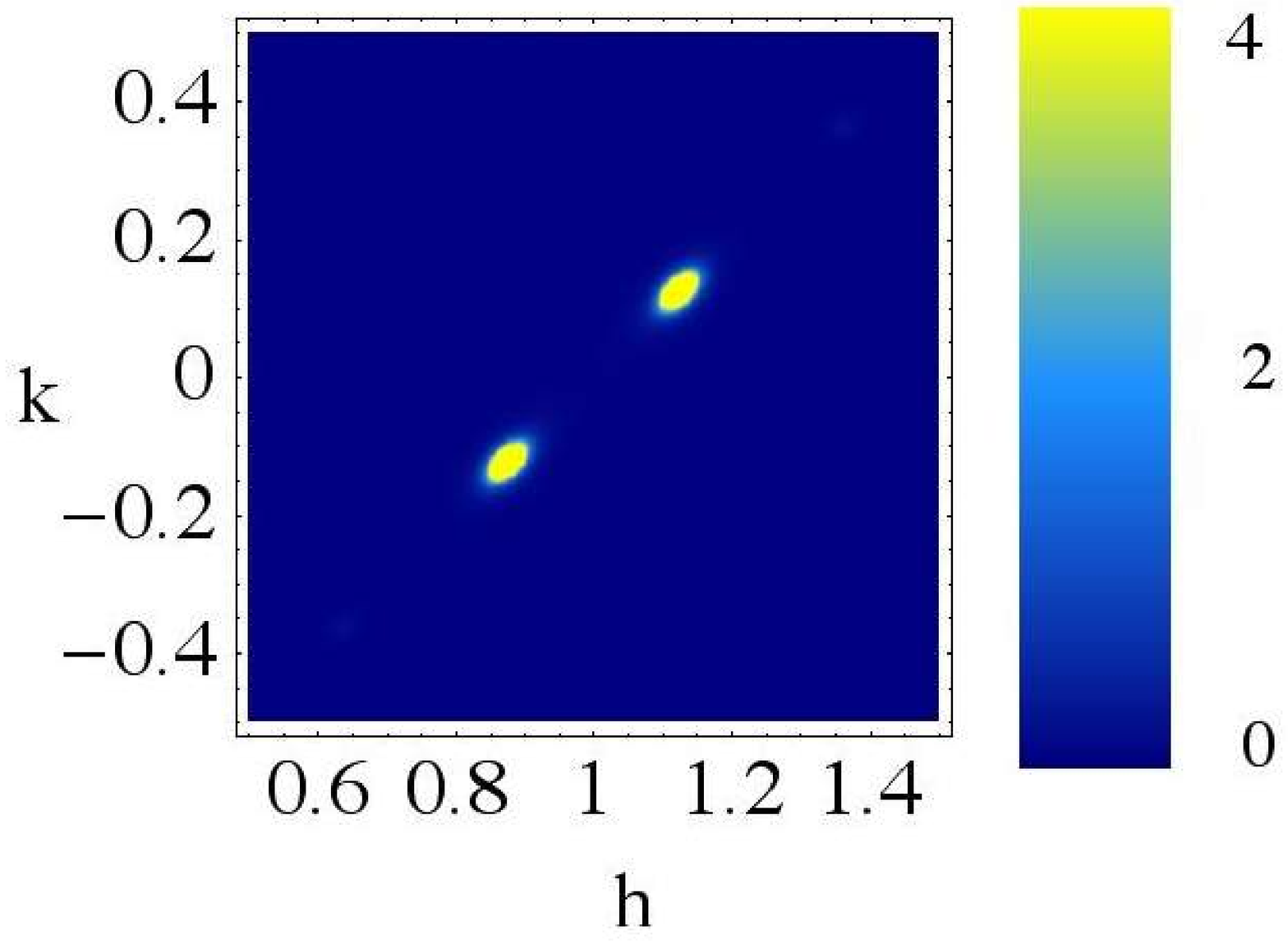}
\epsfysize=.45\textwidth
\includegraphics[height=2.9cm]{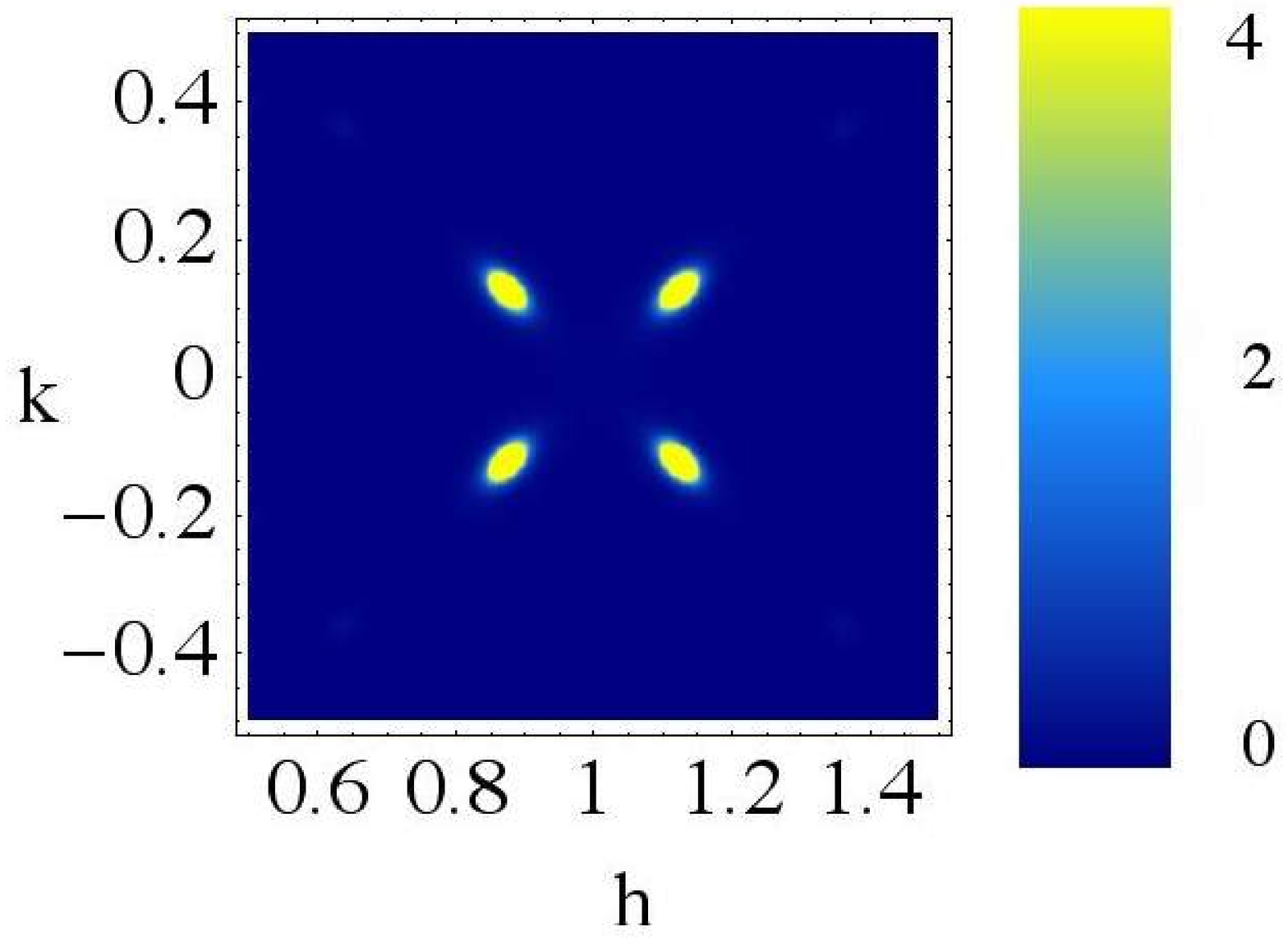}}
\subfigure[$\omega=36meV$]{
\epsfysize=.45\textwidth
\includegraphics[height=2.9cm]{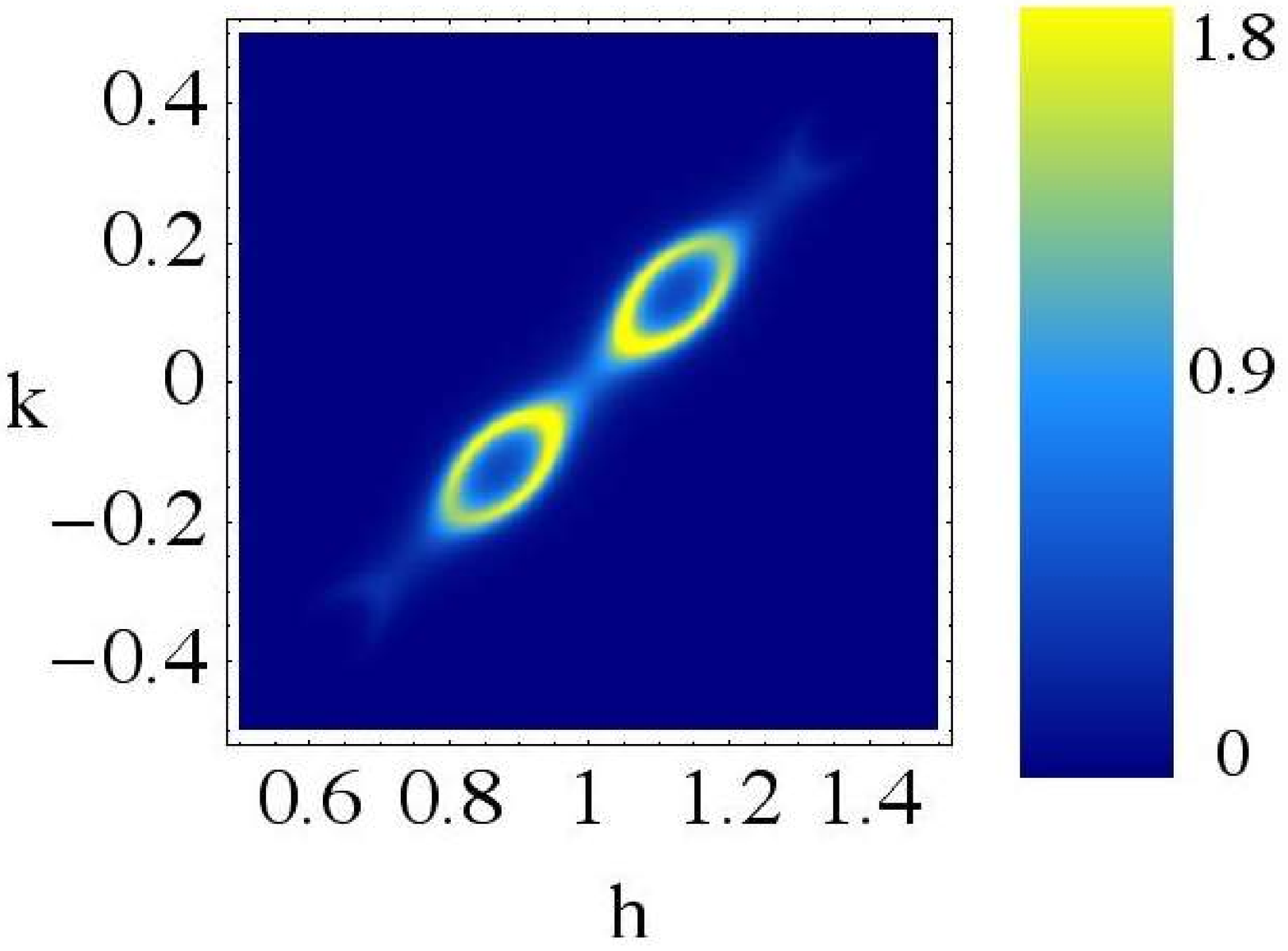}
\epsfysize=.45\textwidth
\includegraphics[height=2.9cm]{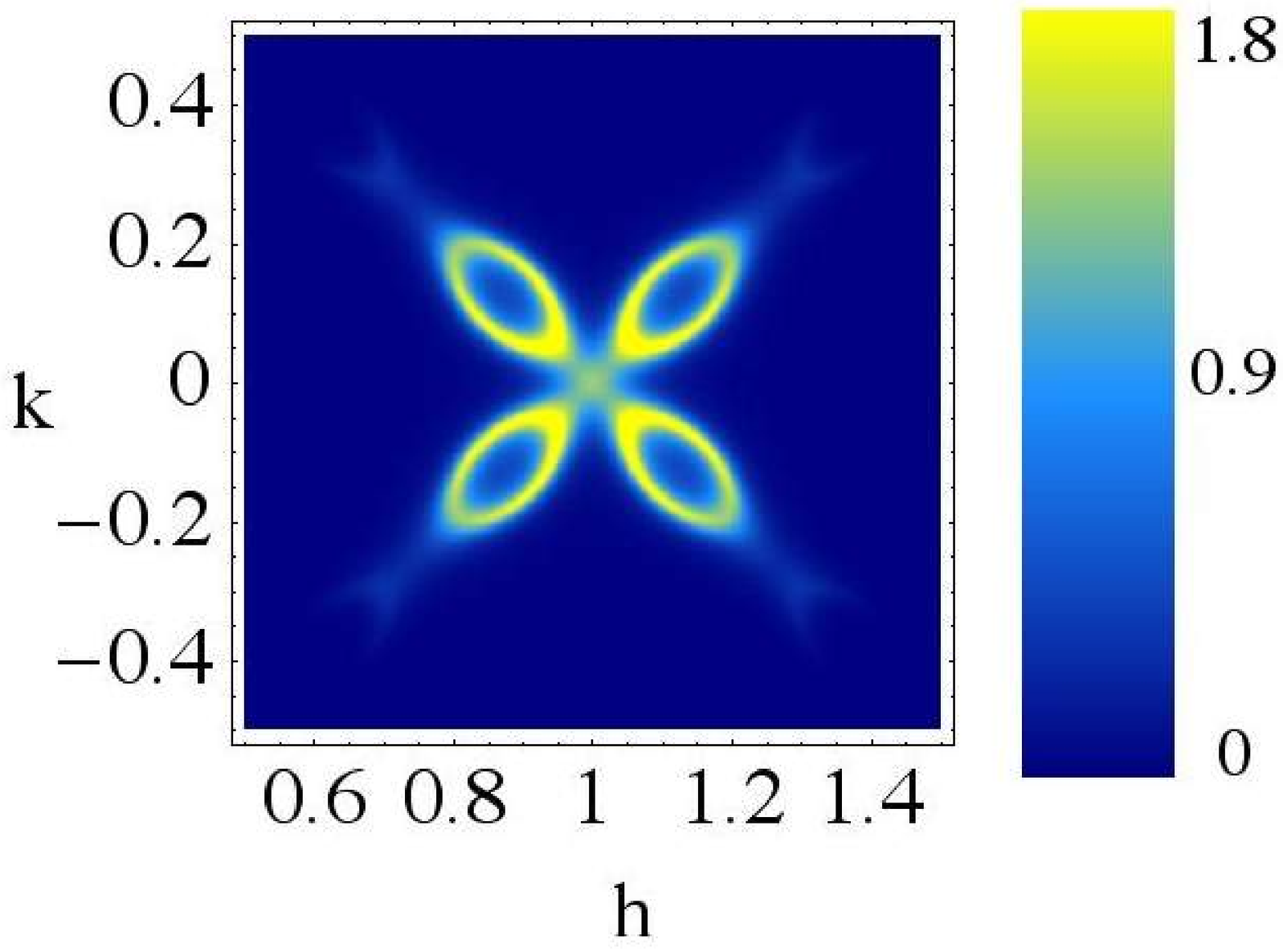}}
\end{center}
\begin{center}
\subfigure[$\omega=55meV$]{
\epsfysize=.45\textwidth
\includegraphics[height=2.85cm]{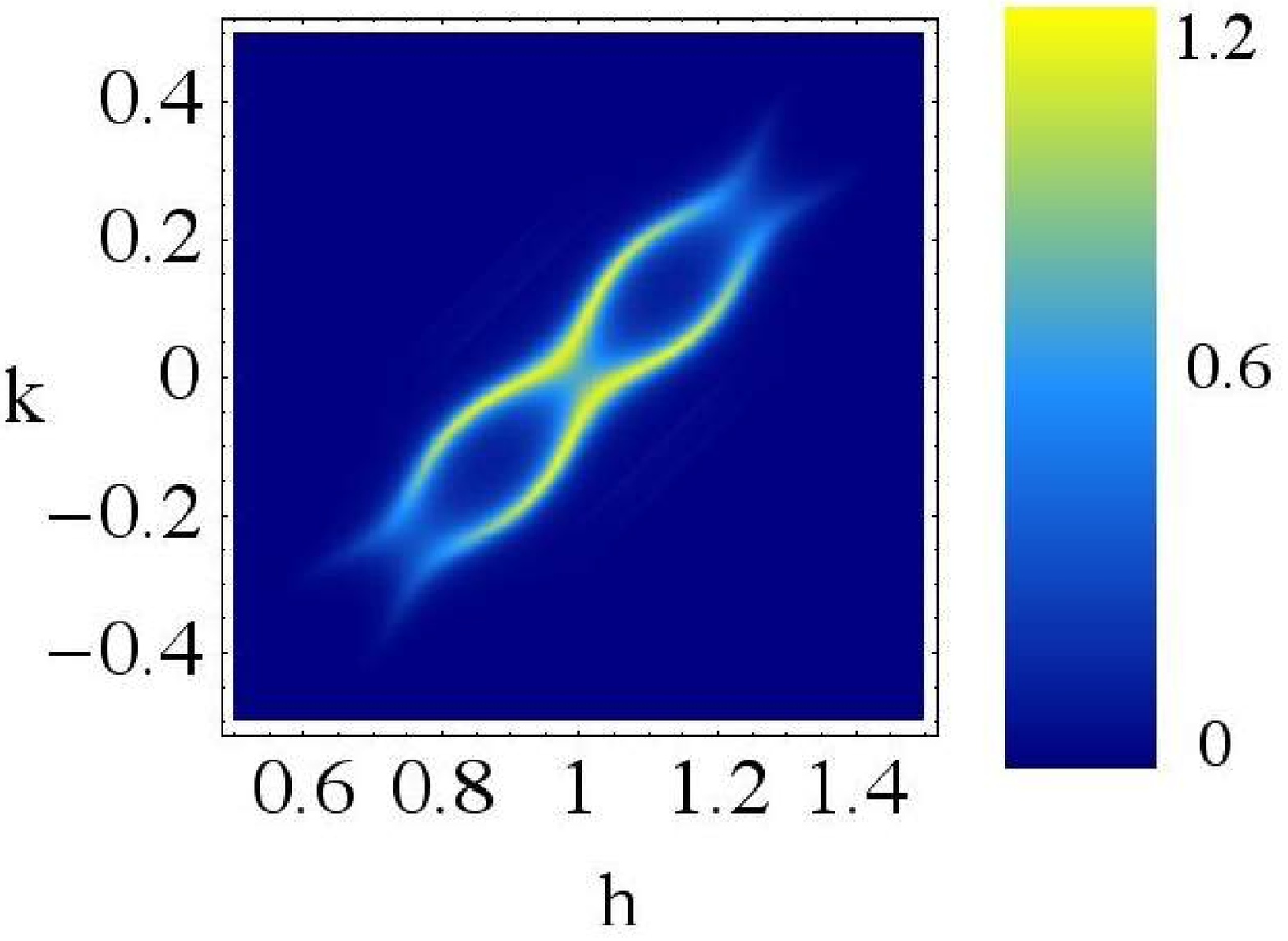}
\epsfysize=.45\textwidth
\includegraphics[height=2.85cm]{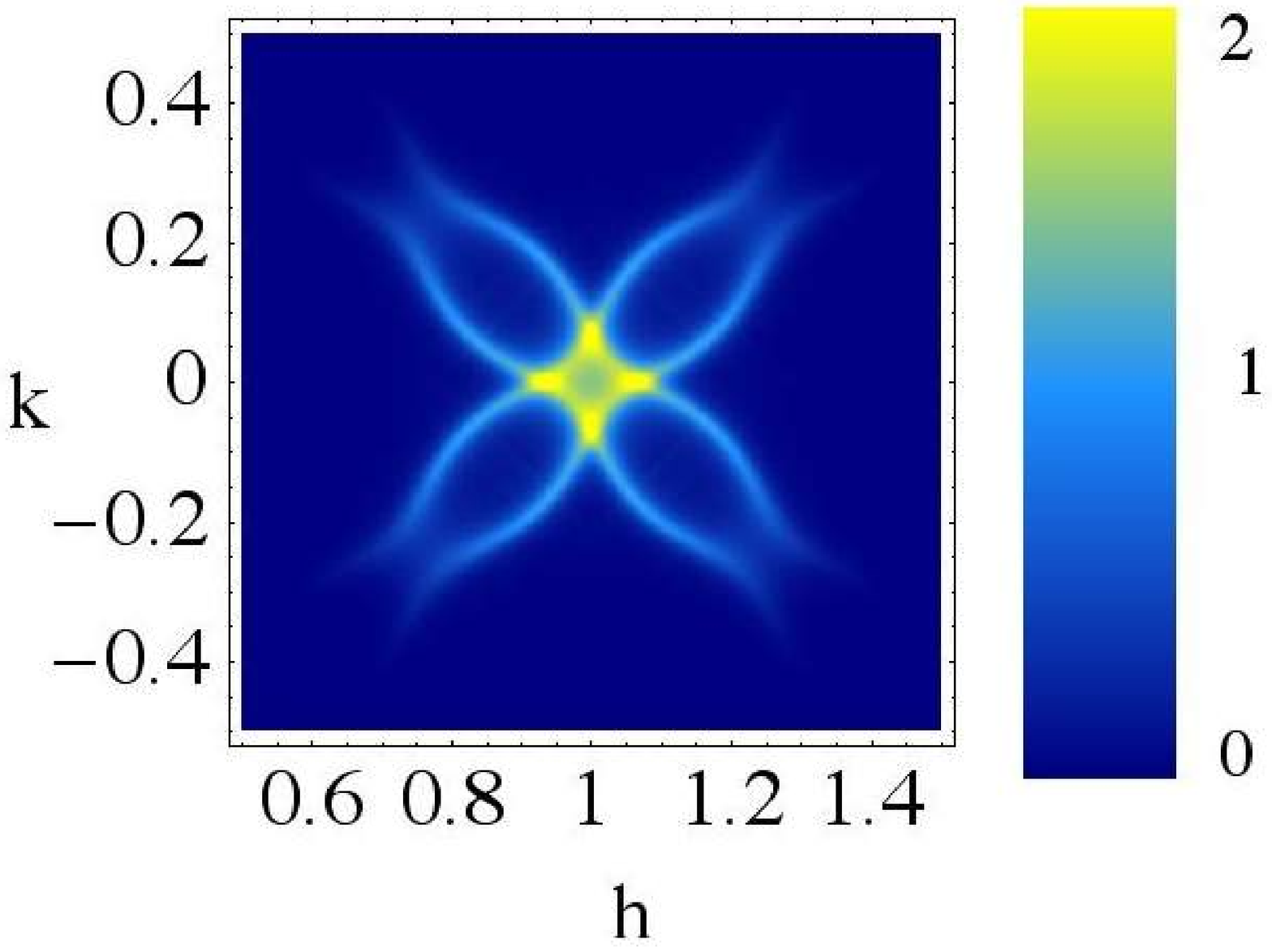}}
\subfigure[$\omega=80meV$]{
\epsfysize=.45\textwidth
\includegraphics[height=2.85cm]{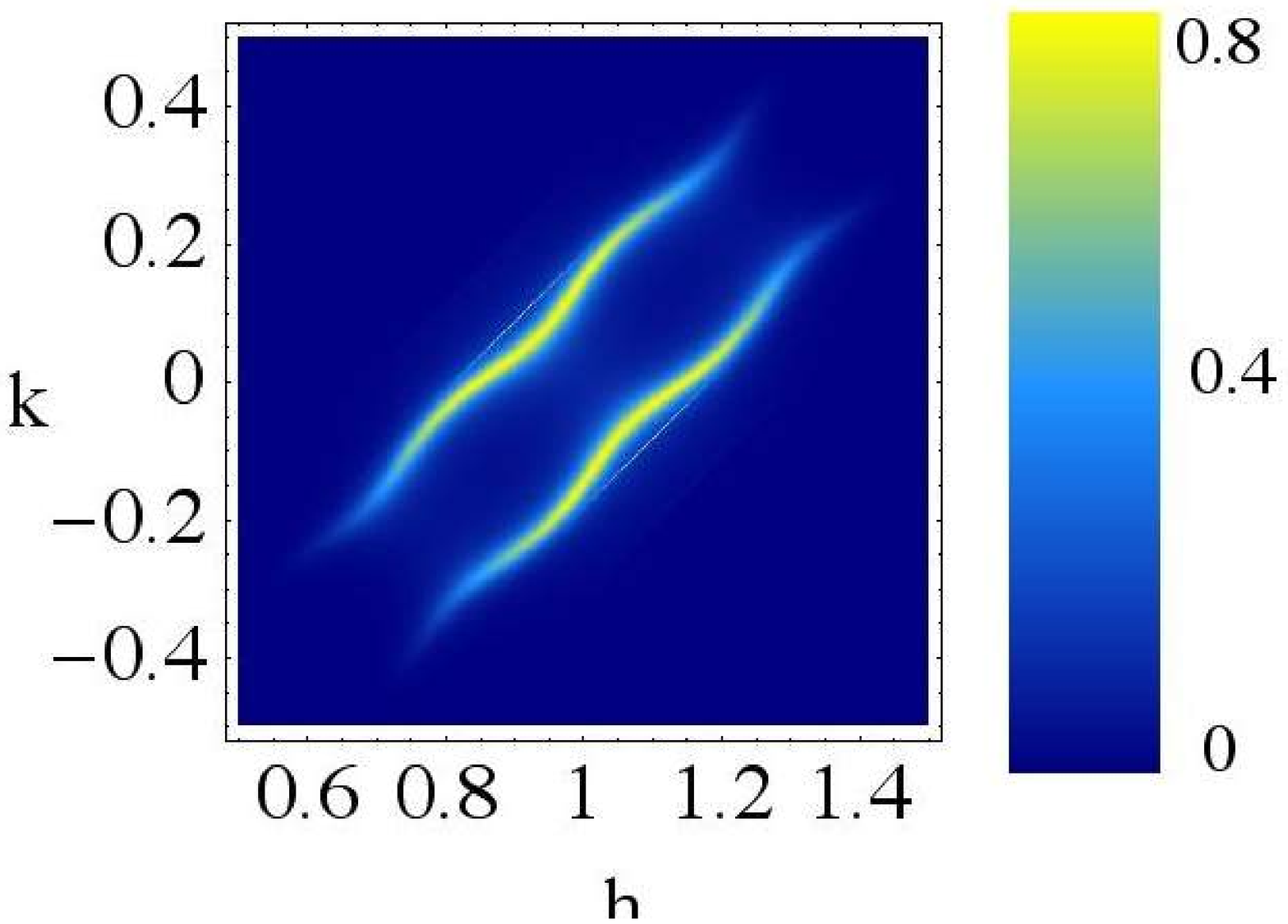}
\epsfysize=.45\textwidth
\includegraphics[height=2.85cm]{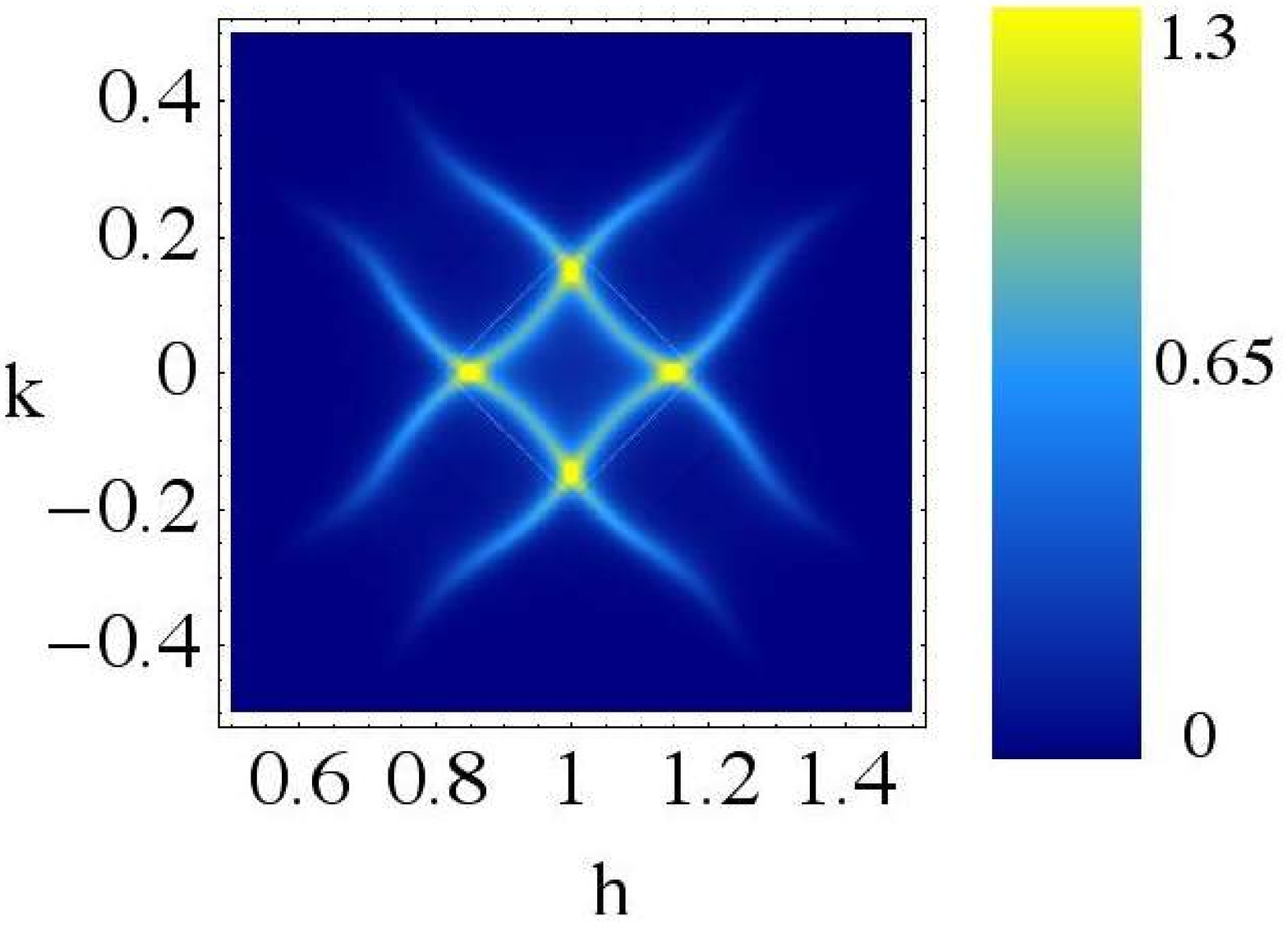}}
\end{center}
\begin{center}
\subfigure[$\omega=120meV$]{
\epsfysize=.45\textwidth
\includegraphics[height=2.85cm]{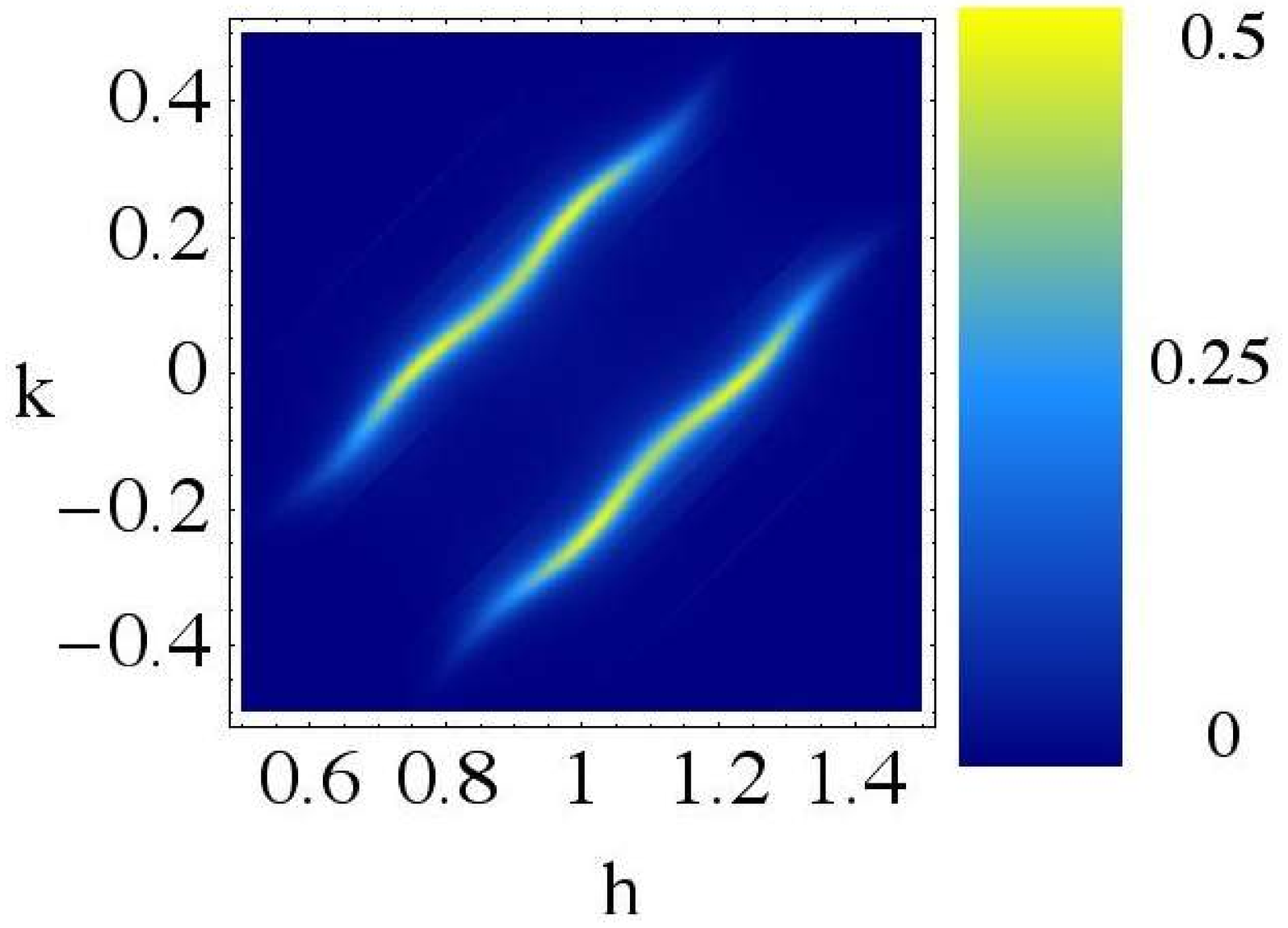}
\epsfysize=.45\textwidth
\includegraphics[height=2.85cm]{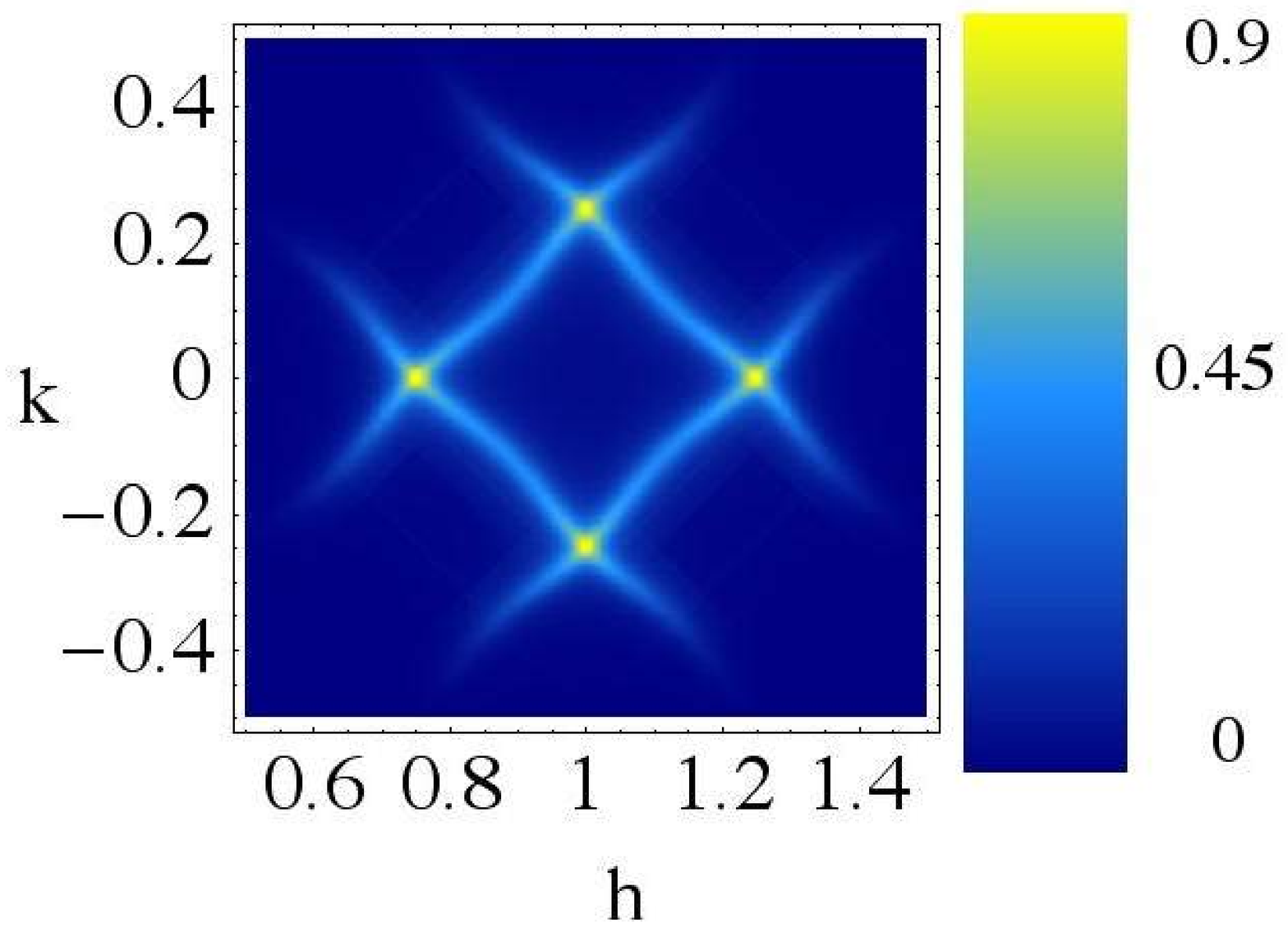}}
\subfigure[$\omega=160meV$]{
\epsfysize=.45\textwidth
\includegraphics[height=2.85cm]{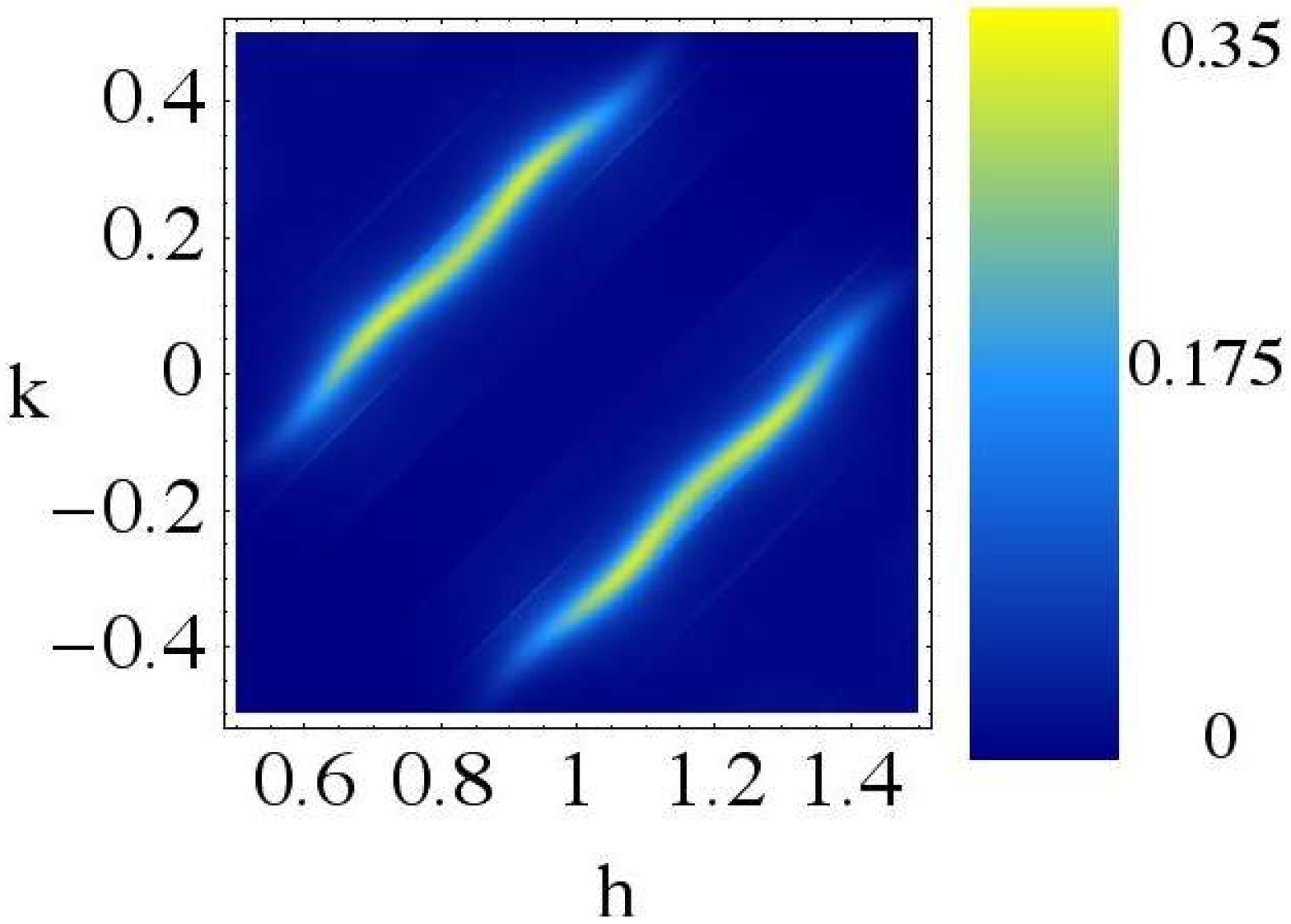}
\epsfysize=.45\textwidth
\includegraphics[height=2.85cm]{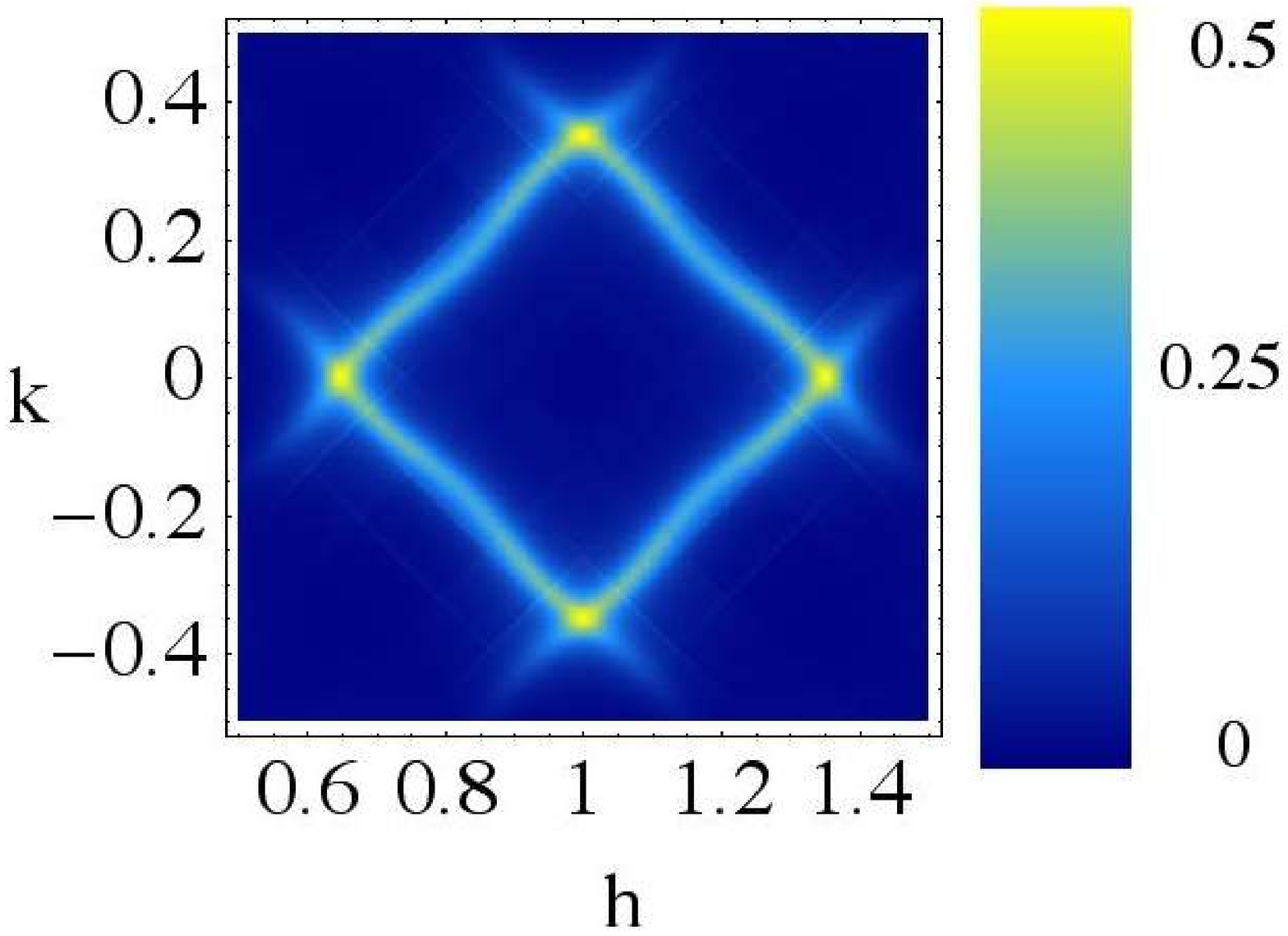}}
\end{center}
\caption{Plots of the scattering intensity in ordering Scenario I 
as a function of $h$ and $k$ (reduced lattice units)
for a number of energies.  At each energy the response is presented
for both a single ladder array (left hand figure) and two ladders arrays orientated
at $90^o$ relative to one another (right hand figure).  The parameters employed here
are discussed in Appendix A1.}
\end{figure*}

\renewcommand{\thesubfigure}{}
\renewcommand{\subfigcapskip}{-4mm}
\begin{figure*}
\begin{center}
\subfigure[$\omega=6meV$]{
\epsfysize=.45\textwidth
\includegraphics[height=2.9cm]{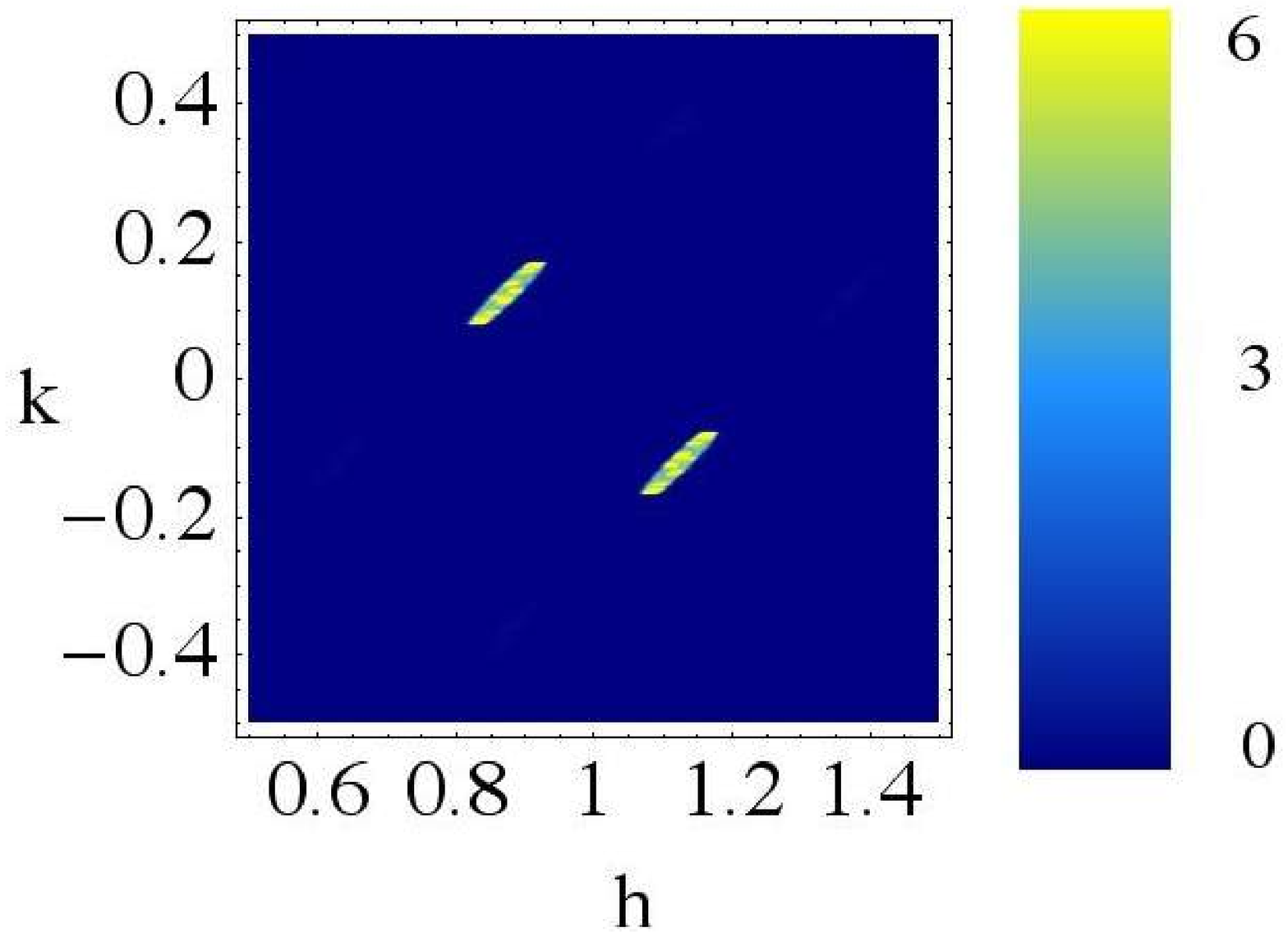}
\epsfysize=.45\textwidth
\includegraphics[height=2.9cm]{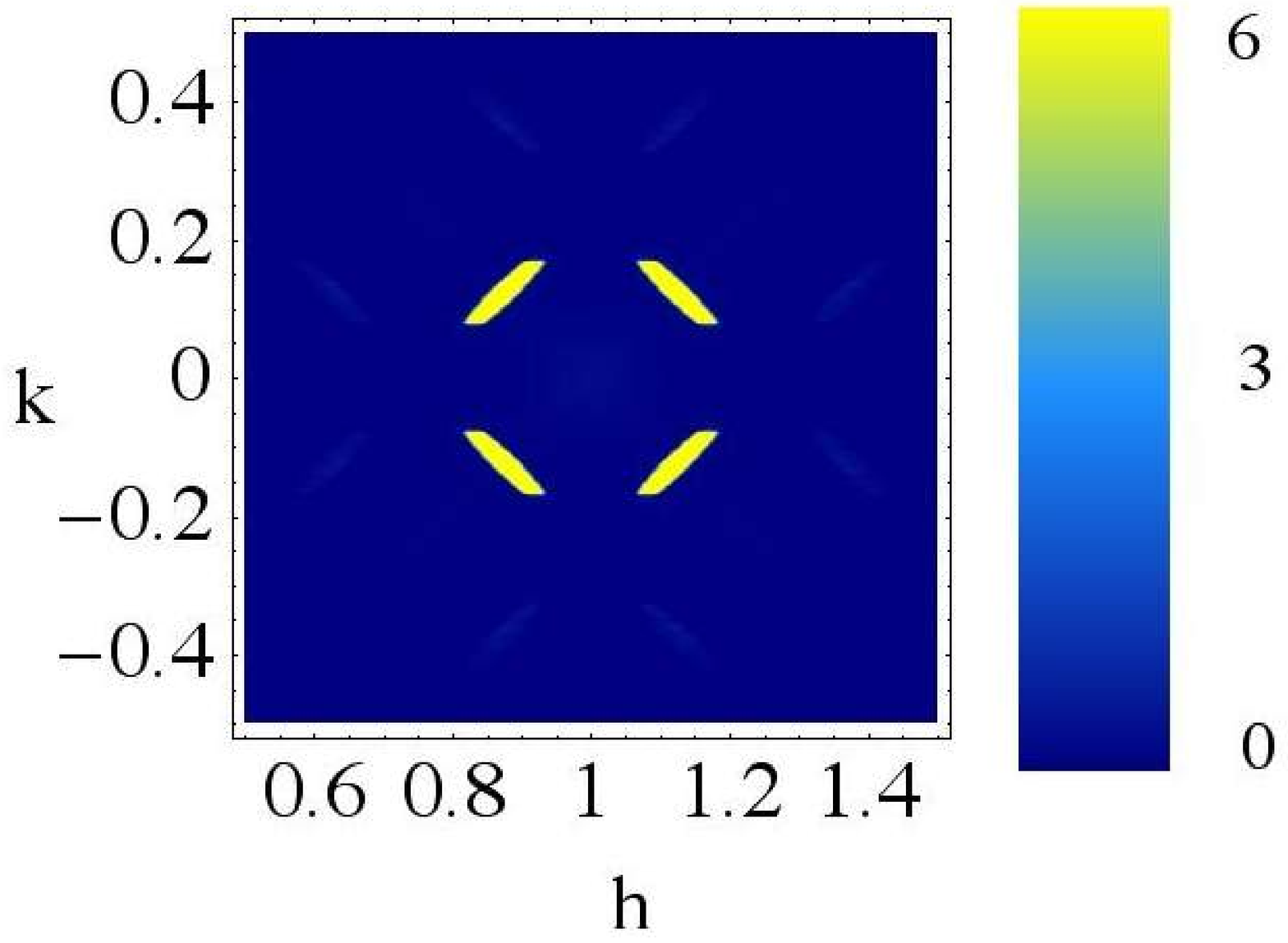}}
\subfigure[$\omega=36meV$]{
\epsfysize=.45\textwidth
\includegraphics[height=2.9cm]{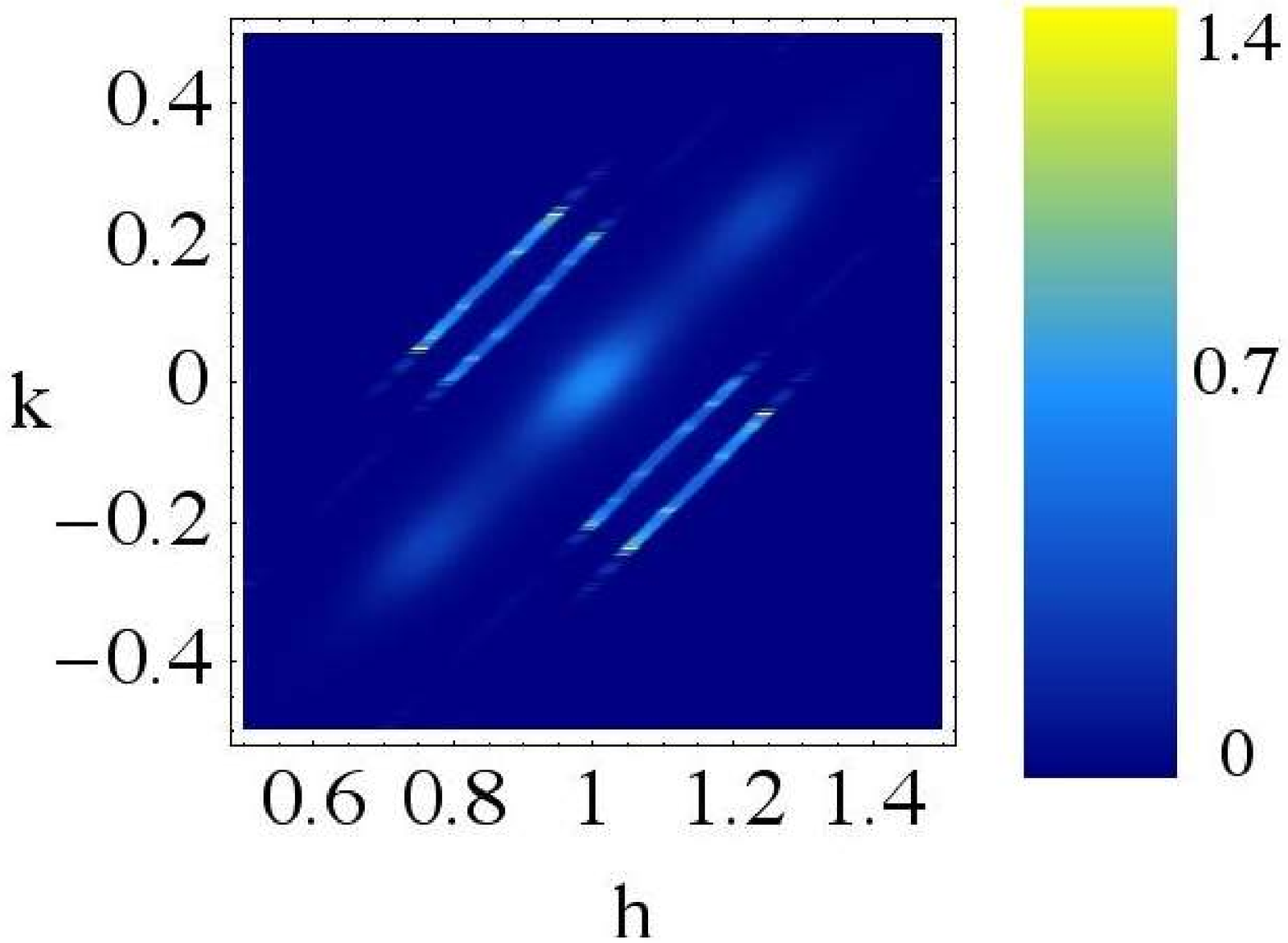}
\epsfysize=.45\textwidth
\includegraphics[height=2.9cm]{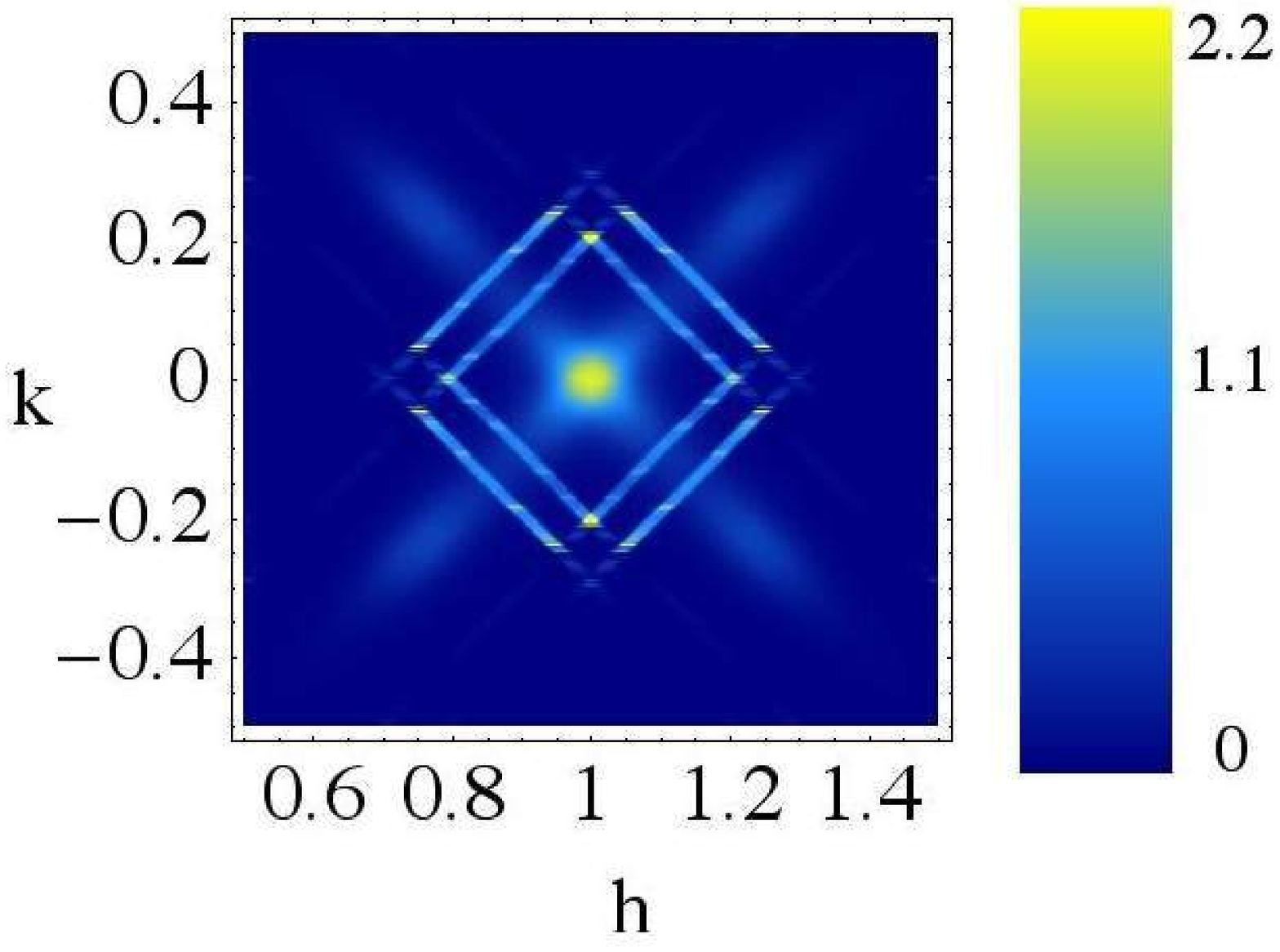}}
\end{center}
\begin{center}
\subfigure[$\omega=55meV$]{
\epsfysize=.45\textwidth
\includegraphics[height=2.85cm]{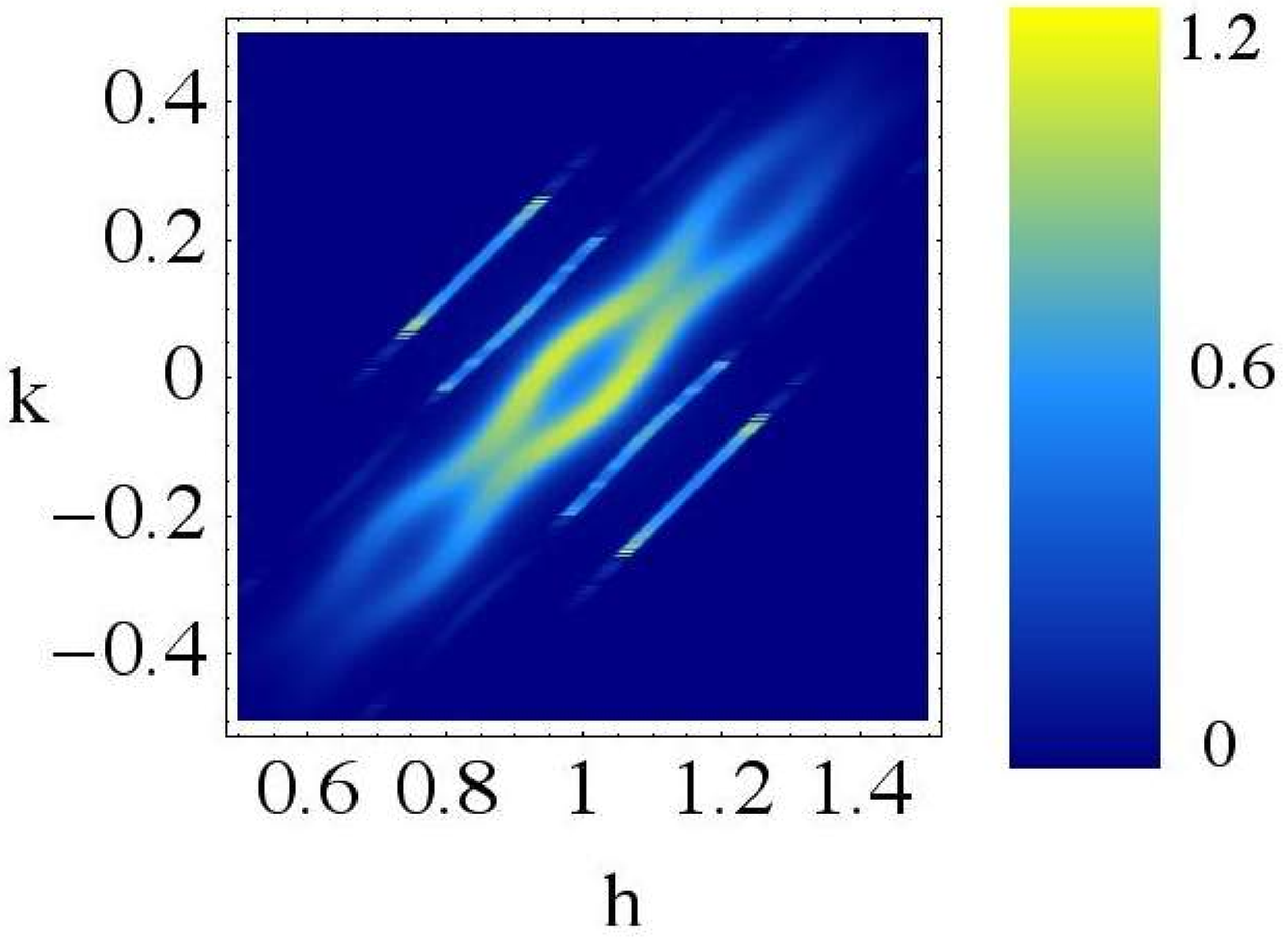}
\epsfysize=.45\textwidth
\includegraphics[height=2.85cm]{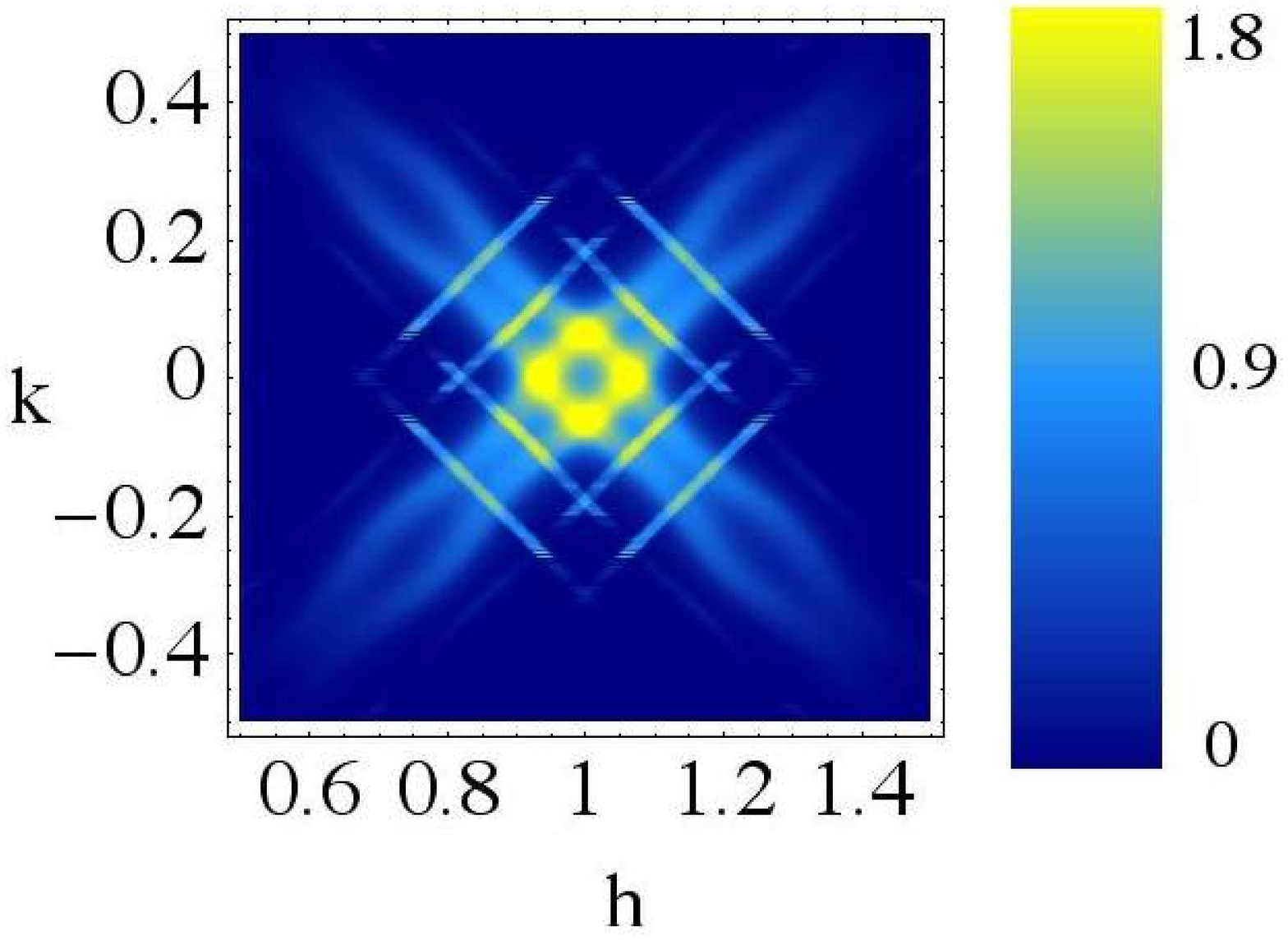}}
\subfigure[$\omega=80meV$]{
\epsfysize=.45\textwidth
\includegraphics[height=2.85cm]{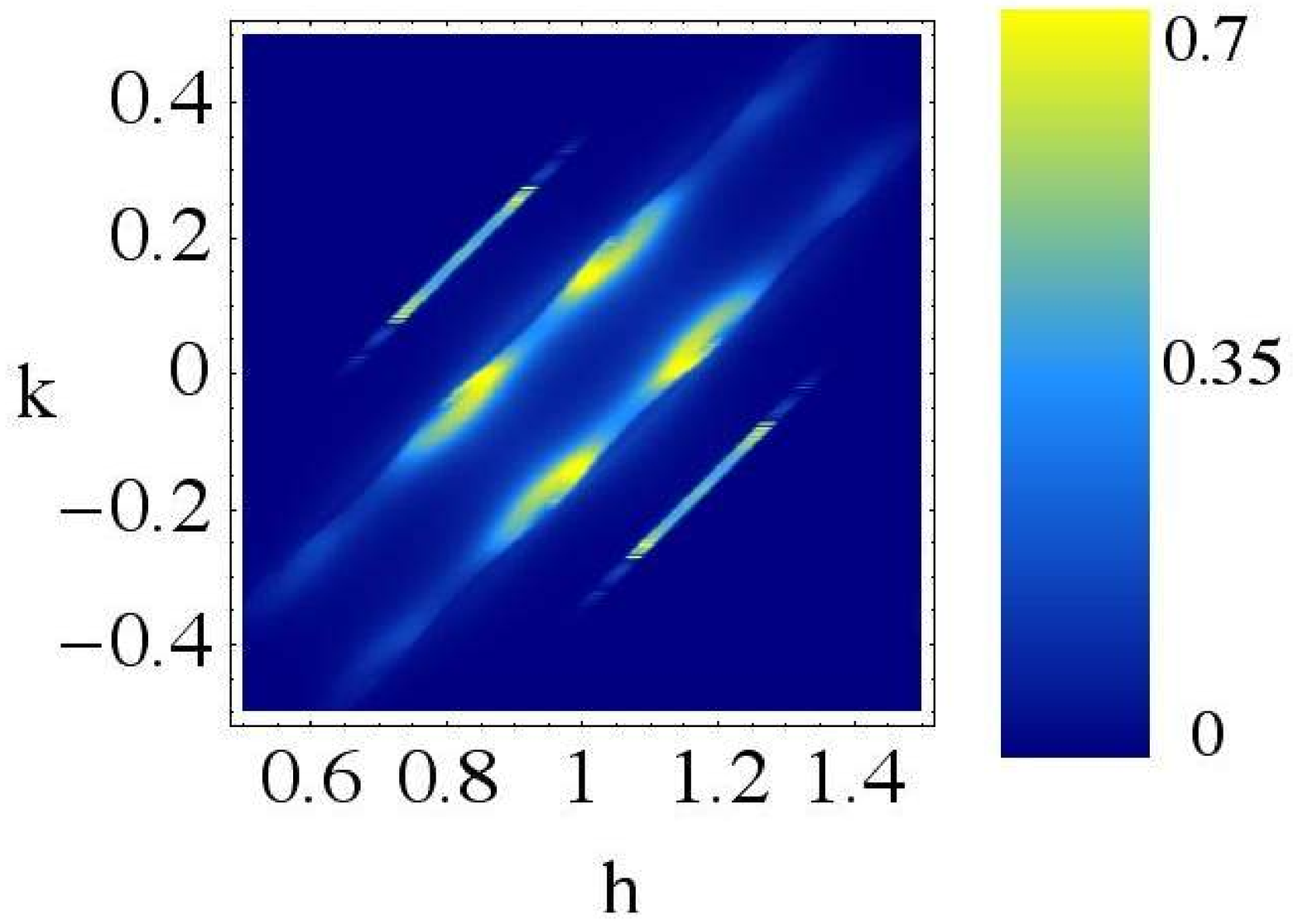}
\epsfysize=.45\textwidth
\includegraphics[height=2.85cm]{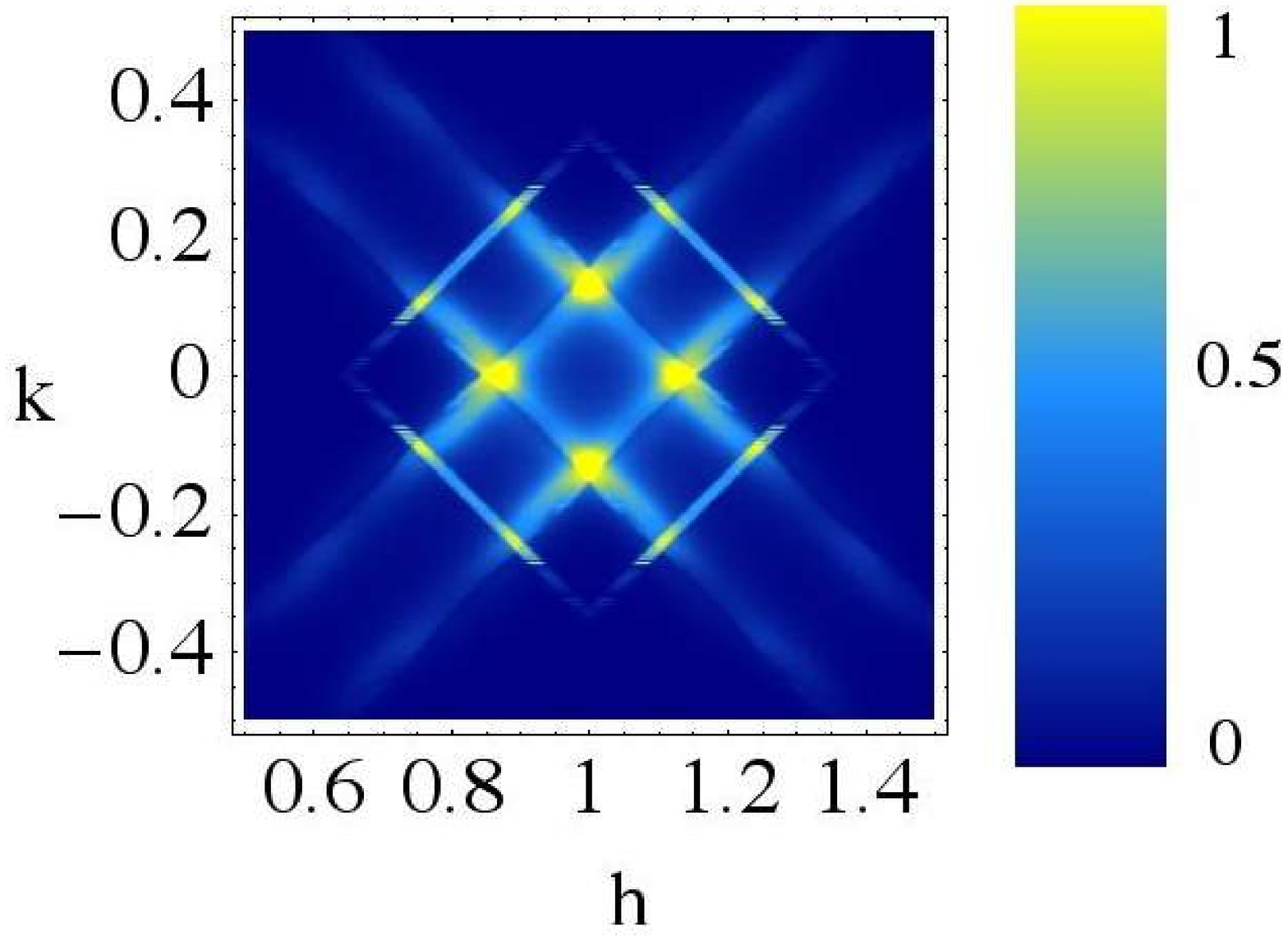}}
\end{center}
\begin{center}
\subfigure[$\omega=120meV$]{
\epsfysize=.45\textwidth
\includegraphics[height=2.85cm]{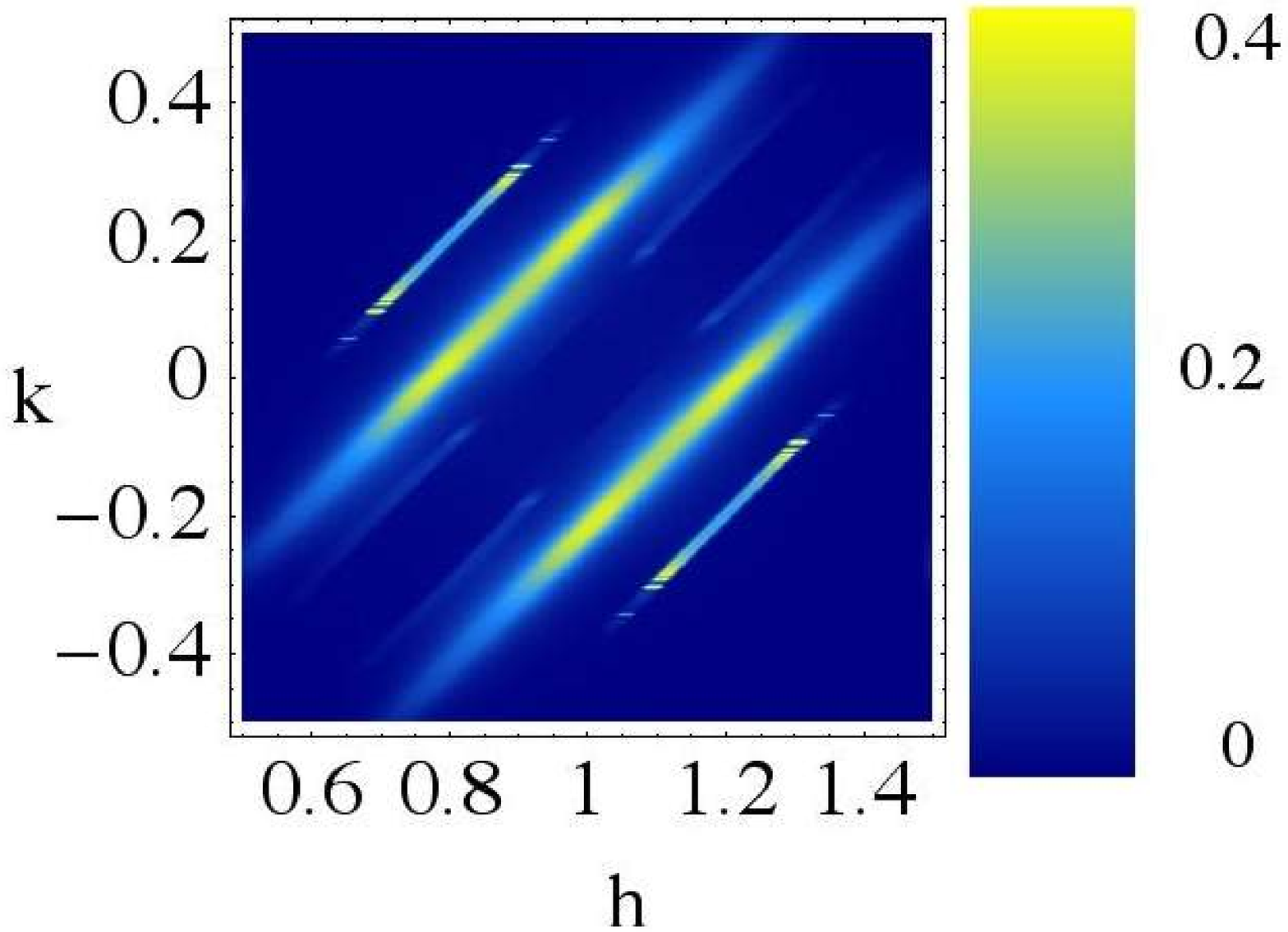}
\epsfysize=.45\textwidth
\includegraphics[height=2.85cm]{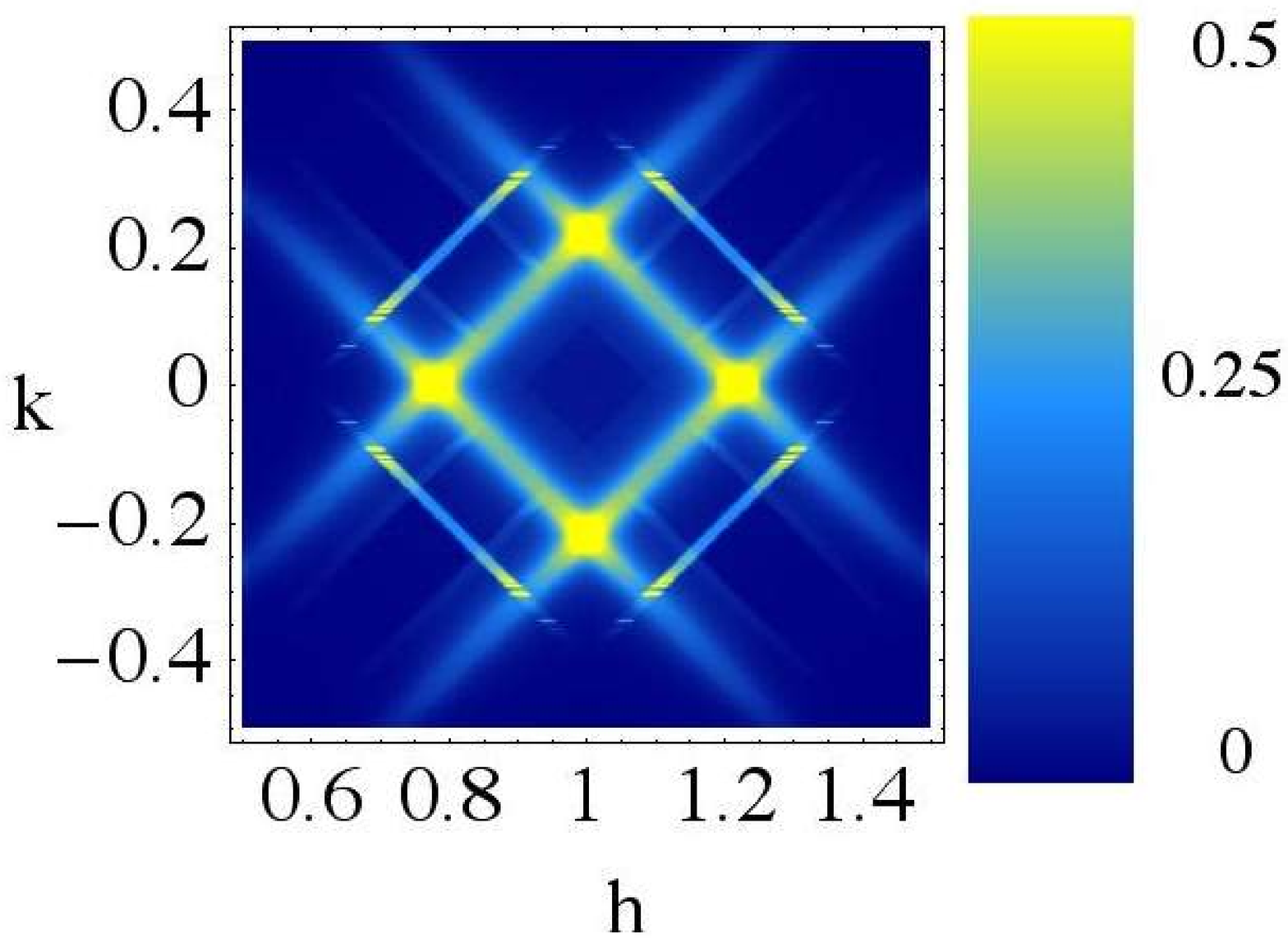}}
\subfigure[$\omega=160meV$]{
\epsfysize=.45\textwidth
\includegraphics[height=2.85cm]{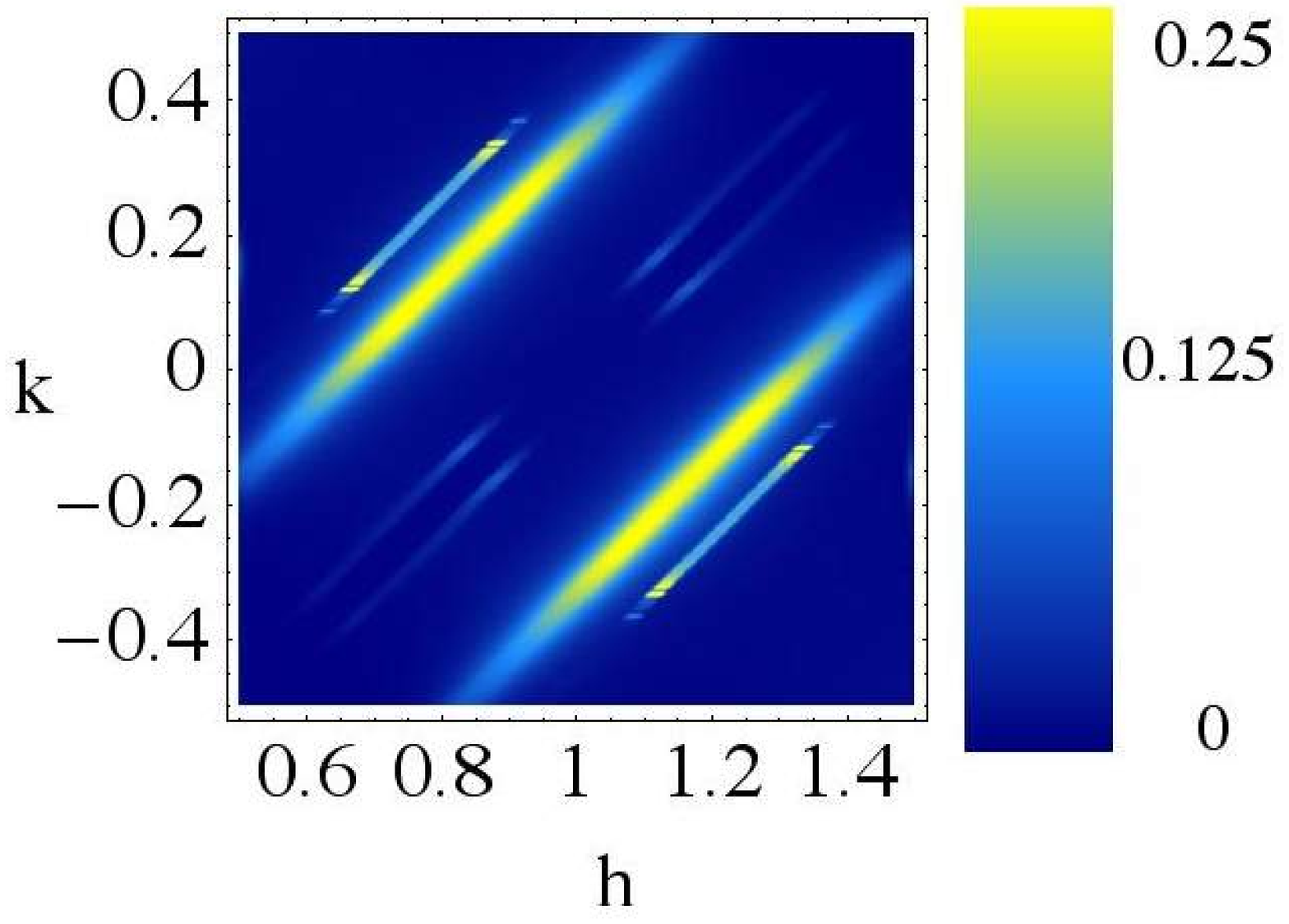}
\epsfysize=.45\textwidth
\includegraphics[height=2.85cm]{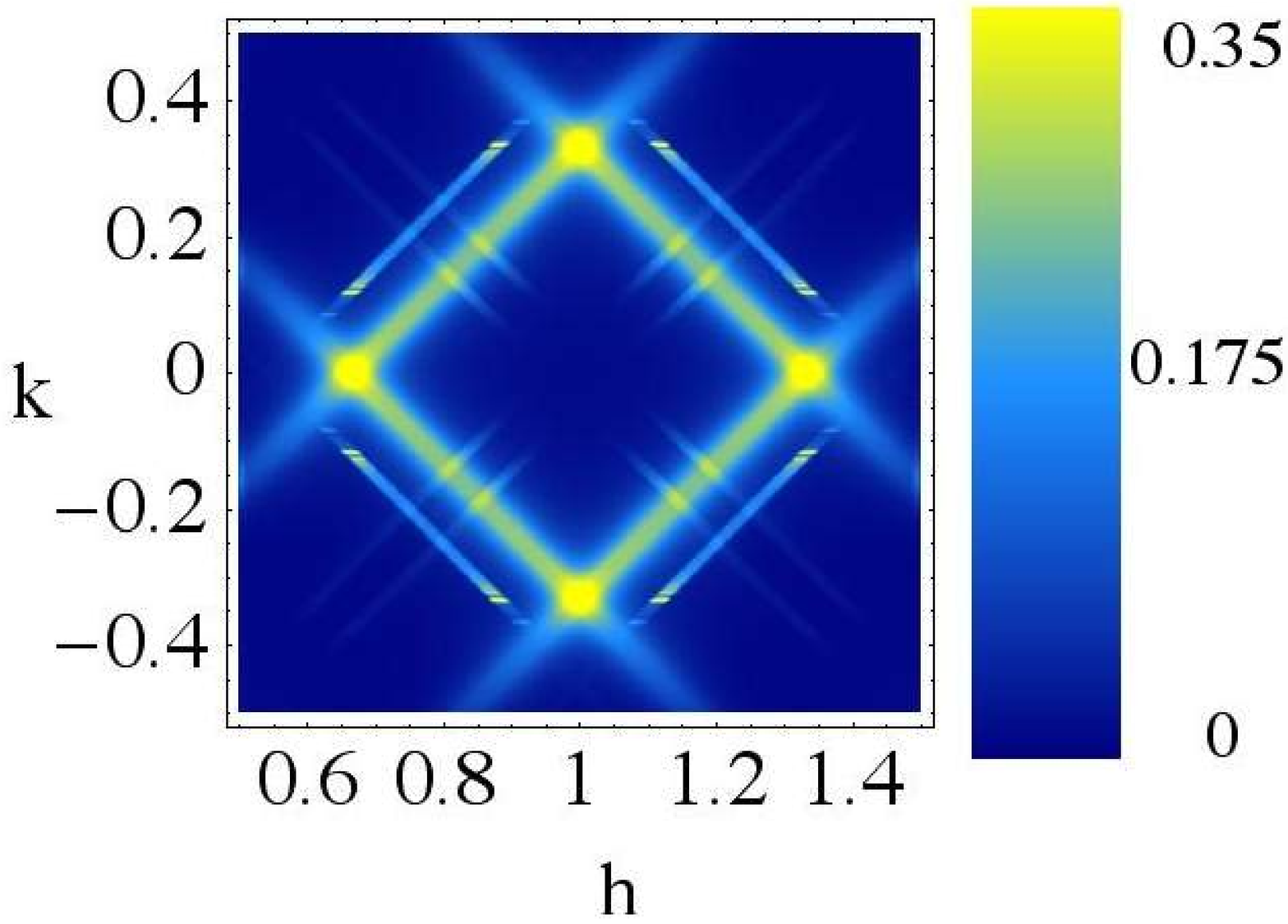}}
\end{center}
\caption{Plots of the scattering intensity in ordering Scenario II 
as a function of $h$ and $k$ (reduced lattice units)
for a number of energies.  The presentation scheme is the same as Fig. 5.
The parameters for the ladders used here are discussed in Appendix A2.}
\end{figure*}

\section{Magnetic Response of Coupled Ladders}

In this section we elaborate upon the magnetic response of the coupled ladders in the first two scenarios
presented in Section IV C.  In particular we will show that either of
these scenarios is compatible with the observed gross features of the magnetic response of $\rm La_{1.875}Ba_{0.125}CuO_4$
and $\rm La_{2-x}Sr_{x}CuO_4$. 

We first consider constant energy slices of the spin response as a function of wavevector.  Choosing the same
energies reported in Ref. \cite{tran_nat}, we plot the results in Figures 4 and 5, where the reduced lattice
units $h$ and $k$ are defined via \cite{tran_nat}
\begin{equation}\label{eIVi}
h = \frac{q_x + q_y}{2\pi};~~~
k = \frac{q_y - q_x}{2\pi}.
\end{equation}
In both figures we show the spin response resulting for a single plane of ladders (left figure of each pair) and 
for a pair of planes of ladders orientated at $90^o$ degrees to one another (right figure of each pair).
The second arrangement (pictured in Fig. 3) corresponds to how stripes
order in $\rm La_{1.875}Ba_{0.125}CuO_4$ 
and $\rm La_{1.82}Sr_{0.18}CuO_4$, and so is the one relevant for
comparison with experiment. We, however, include the response of a
single plane as it is here that the magnetic response of the two
ordering scenarios most sharply distinguish themselves. 

In Figure 5, Scenario I is presented, the
case where the magnetic order develops perpendicular to the ladder, i.e. at wavevector $(\pi,\pi\pm \frac{\pi}{4})$.
At the lowest of energies shown, $\omega=6meV$, we
find for a single plane of ladders,
a pair of incommensurate spin waves dispersing at the incommensurate wavevectors $(h,k)=(1\pm 1/8,\pm 1/8)$.
In the response for two planes, we then observe a second pair of spin waves, rotated by $90^o$ relative
to the first.
The dispersions cones of the wavevectors are elongated along the diagonal, a consequence of the anisotropy
between inter- and intra-ladder couplings ($J_c$ and $J$).
As we increase in energy to $\omega=36meV$, the spin waves disperse outwards.  We see that spectral weight
of the cones in anisotropically distributed, with more weight being found on the side of the cone
nearest to $(\pi,\pi)$.  Thus we see that as we increase in energy, the spectral weight
appears to move away from the incommensurate points towards $(\pi,\pi)$.
By $\omega = 55meV$, the energy corresponding to the gap in the undoped ladders, 
the cones have begun to overlap.  This overlap is enhanced 
for the response of a pair of planes, leading to the most intense response coming from $(\pi,\pi)$.
As energy is further increased, we observe a rotation in the intensity 
by $45^o$ in the pair plane response (compare energies $\omega=36meV$ and
$\omega=80meV$).   The rotation results from the dominance of
the spin response of the half-filled ladders, which for a single plane form lines of intensity.  With a 
pair of relatively orientated planes, the lines cross, leading to the four peaks.
As we increase energy further, the peaks in the two-plane response
disperse outwards while at the same time losing intensity.  By
$\omega=160meV$, the spin response has become both comparatively broad and weak.

In Figure 6, we plot the magnetic response of the second scenario where order appears parallel
to the ladders.  At the lowest energy shown, $\omega = 6meV$, we again find incommensurate spin waves,
which now appear, for a single plane, at $(h,k)=(1\mp 1/8,\pm 1/8)$.  The spin waves, in this case,
are much more strongly anisotropic.  But we believe this is a feature of the details of the ladder susceptibilities,
not a fundamental feature of the model.  While the response at $\omega = 6meV$ for this
ordering pattern is rotated by $90^0$
relative to where the order develops perpendicular to the ladders, the response for a pair
of planes is qualitatively no different in the two cases.  
As we increase in energy, the spin waves appearing at low energies evolve
into the response of a set of nearly uncoupled doped ladders.  Accompanying this
evolution is a separate development of spectral weight at $(\pi,\pi)$.  The presence
of inelastic spectral weight of $(\pi,\pi)$ is a consequence of the competition in this scenario between
order developing at $q_x = \pi$ (the location of coherent mode on the doped ladder)
and order developing incommensurately
at $q_x = 3\pi/4$ and $5\pi/4$ (the location of the quasi-coherent mode on the undoped ladder).  
Though in this case order is favored at
the incommensurate wavevector, the commensurate order remains pre-emergent and so
appears at finite energy values.
Taken together, these two effects again give the appearance of a movement of spectral weight
towards $(\pi,\pi)$.  As we continue further up in energy, the response of the half-filled ladders
again begins to dominate, with a peak in the intensity near $(\pi,\pi)$ (see $\omega = 55meV$ in Fig. 6).  
The dominance of the half-filled ladders then continues to higher energies ($\omega=80meV$ and above), and
consequently, the response takes on the same form as that of Figure 5.

\begin{figure}
\begin{center}
\noindent
\includegraphics[height=8cm,width=10cm,angle=0]{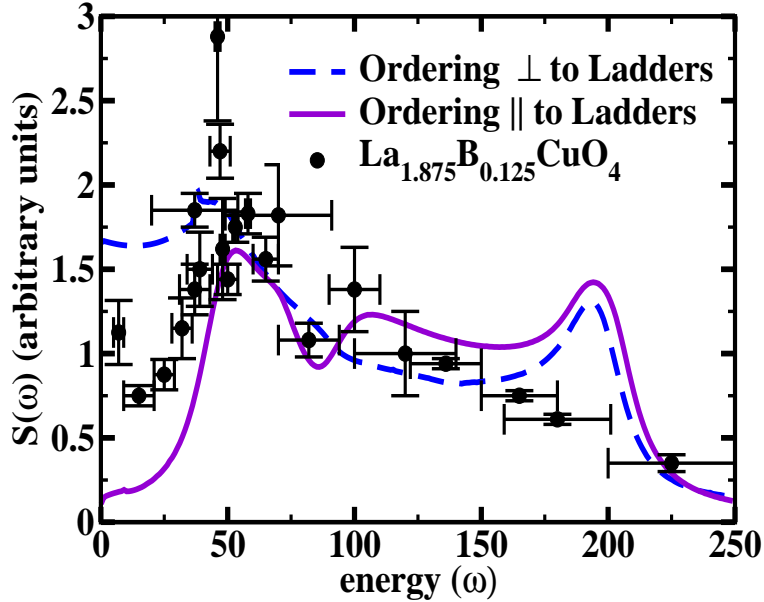}
\end{center}
\caption{Integrated intensity, $S(\omega )$, of the coupled ladder
system in the two ordering scenarios.  For comparison, we plot these
results against Ref. \cite{tran_nat}.  The parameters are the same as
for Figs. 5 and 6.}
\end{figure}

While our model of coupled ladders finds reasonable approximates to
the observations on LBCO of Ref. \onlinecite{tran_nat}, it produces features
at higher energies that are typically much sharper than those actually observed.
This however is not surprising.  The RPA approximation we employ will generically
underdamp high energy stripe-like correlations.

As another measure of the response of the two scenarios, we 
compute at fixed energy the $q$-integrated intensity, $S(\omega)$, of the coupled ladders.
This quantity is defined by
\begin{equation}\label{eIVii}
S(\omega) = \int d^2q {\rm Im}\chi (\omega,q_x,q_y).
\end{equation}
We plot the results in Figure 5 for the two scenarios.
We see we obtain a rough agreement.  At low energies there is an increase in
intensity corresponding to the development of incommensurate long range order.  We also see
an enhancement in the intensity at $J=50meV$ which corresponds to the spin gap of the half-filled
ladders.  This is to be expected as the excitation spectrum of the half-filled ladder is a single coherent mode
and should have a strong response.  For energies in excess of the spin gap, we then see a gradual decline
in intensity in both the measured and computed responses.  The primary difference between the two
scenarios lies in the total amount of spectral weight found at low energies.  But this difference
is not fundamental and rather is a product of particular choices made to describe the susceptibilities
of the individual ladders in both cases.

\section{Discussion}

We approached this work with three primary motivations:
i) to understand whether a model of coupled ladders with alternating
levels of doping is compatible with the observed spin
response in $\rm La_{1.875}Ba_{0.125}CuO_4$ and $\rm La_{2-x}Sr_xCuO_4$;  
ii) to suggest that distinct patterns of magnetic
ordering can lead to the same set  of experimental observations in
these particular cuprates; and iii) to explore the specific role that
the internal  dynamics of the doped regions play in the development of
stripe order. We deal with each in turn.

The observations of Ref. \cite{tran_nat} of the inelastic
spin response in $La_{1.875}Ba_{0.125}CuO_4$ has several basic features.
At zero energy there appear inelastic incommensurate spin waves at the
four wavevectors, $Q = (\pi ,\pi(1\pm 1/4))$ and $Q = (\pi(1\pm 1/4),\pi )$.
At small but finite energies (up to $50 meV$), the intensity associated with these spin waves appears
to propagate away from these incommensurate points and inwards
towards $(\pi,\pi)$.  At higher energies, the movement of the
spectral weight reverses direction, propagating outward, but with peaks rotated by
$45^o$ relative to the low energy spin waves. 

In the previous sections we have show that there exist two coupled ladder scenarios
that qualitatively reproduce these features.
The two scenarios are distinguished by the direction in which the incommensurate order
develops, either parallel or perpendicular to the direction of the
ladder.  Nonetheless in both scenarios we find magnetic order at the
four incommensurate wavevectors (in bi-planar systems).  In both, 
we have a movement of spectral weight towards $(\pi,\pi)$ as energy moves upwards
to $50 meV$.  And for energies greater than $50meV$, the spin response of both scenarios, dominated
by the coherent mode of the undoped ladder, yields the same outward (from $(\pi,\pi)$) propagation
of spectral weight rotated by $45^o$ relative to the location of the low energy spin waves.

While the two scenarios are qualitatively similar, there are
quantitative differences. 
The low energy spectral features present in Scenario II (ordering along the ladder) are
far more anisotropic than those in Scenario I (ordering perpendicular to the ladder).  This 
however is less a fundamental feature of Scenario II and more a consequence of the use
of the field theoretic treatment of the doped ladders at medium energies.  At such energies,
the neglected
non-relativistic band curvature will moderate anisotropic features.  A more fundamental difference
is the manner in which spectral weight moves inwards towards $(\pi,\pi)$ as energies are increased
to $50 meV$.  In Scenario I, this movement is a consequence of expanding spin wave cones possessing an unequal
distribution of spectral weight.  In Scenario II, the weight moves towards $(\pi,\pi)$ due to the spectral
features present on the uncoupled doped ladder together with nascent ordering at $(\pi,\pi)$.  This
ordering, discussed briefly in Section IV C, while not elastic, cannot be entirely suppressed at higher
energies.

While the primary focus of this work has been on static
stripe order of the kind observed in $\rm La_{1.875}Ba_{0.125}CuO_4$,  
one may ask whether our model of coupled ladders
might be applicable more generally to incommensurate magnetic
excitations in the cuprates. At least to some degree it does.
In slightly overdoped LSCO ($\rm La_{1.84}Sr_{0.16}CuO_4$), nascent
incommensurate long range order has been reported \cite{lake1,christensen,vignolle}.
Specifically, neutron scattering experiments observe a broad peak centered
about $11meV$ in the scattering intensity at the
incommensurate wavevector $(h,k)=(1\pm 1/8,\pm 1/8)$ (see Figure 8).
(A similar phenomena is seen in $\rm La_{1.82}Sr_{0.18}CuO_4$ \cite{tran_lsco}.)
This magnetic response can be understood through our model of ladders
with an interladder coupling, $J_c$, less than its critical ordering
value, $J_{crit}$.  The responses for $J_c < J_{crit}$ for both
scenarios ($||$ and $\perp$) are pictured in Figure 8.  For comparison
we have plotted the corresponding responses for $J_c=J_{crit}$ where
long range order is fully developed. There, as $\omega$ is decreased,
the response diverges at the incommensurate wavevector. 

\begin{figure}
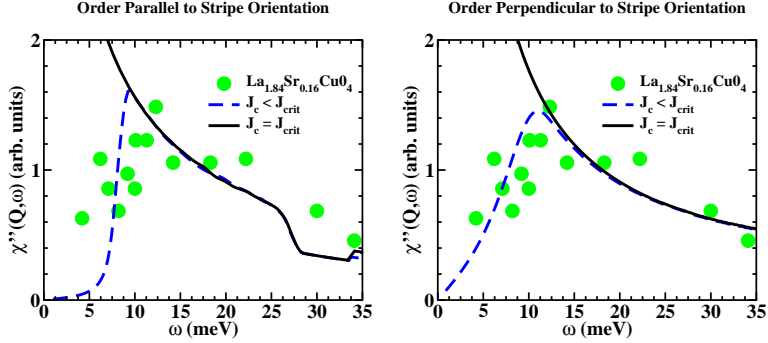

\begin{center}
\noindent
\subfigure[$$]{
\epsfysize=.45\textwidth
\includegraphics[height=4.5cm]{max_par.eps}
\epsfysize=.45\textwidth\hskip .1in
\includegraphics[height=4.5cm]{max_perp.eps}}
\end{center}
\caption{The intensity of the spin response at the incommensurate ordering wavevector, $Q$,
for the two ordering scenarios.
We show the response for both sub-critical and critical values of the
interladder coupling, $J_c$.  We compare this with the maximal
intensity in $\rm La_{1.84}Sr_{0.16}CuO_4$ for $T>T_c$ measured in Ref. \cite{christensen}.
For ordering perpendicular to the ladders the parameters are the same as those
found in Figs. 5 and 7.  For the second scenario, as explained in Appendix A2, the parameters used
differed from those employed in Figs. 6 and 7.} 
\end{figure}

The compatibility of both stripe ordering scenarios we
discuss with the observed spin response in both $\rm La_{1.875}Ba_{0.125}CuO_4$ and 
$\rm La_{1.84}Sr_{0.16}CuO_4$ trades on the bilayer ordering structure of
stripes in these two materials where the stripes in adjacent
copper-oxide planes are orientated perpendicular to one another.
While the spin responses of stripes in a single plane is much
different in the two scenarios, once rotated and superimposed, the
responses become qualitatively the same. 

Experimental evidence beyond the spin response in these compounds 
provides only limited evidence allowing one to distinguish between
these two scenarios. 
In Scenario I extended to general levels of doping \cite{ZEK}, 
incommensurate wave vectors of charge and spin density waves are
arranged perpendicular to the direction of stripes and are related to each
other via
\be\label{ek}
{\bf Q^I}_{ch} = (0,\pm 4\pi x), ~~ {\bf Q^I}_{s} = (\pi,\pi \pm 2\pi x) ,
\ee
where $x$ denotes the doping.
In contrast, in Scenario II stripes are found in the form of coupled two
leg ladders, independent of doping, gaining in this fashion, magnetic energy.
Here the pattern of incommensuration appears as
\be\label{we}
{\bf Q^{II}}_{ch} = (0,\pm \frac{\pi}{2}), ~~~ {\bf Q^{II}}_{s} = (\pi \pm 2\pi x,\pi) .
\ee
The behavior of incommensurate spin order at $Q_{sx} = \pi \pm 2\pi x$ as a function of doping
directly tracks the wavevector where low energy quasi-coherent modes exist in the doped ladder.

To distinguish between these two scenarios we must then focus upon the charge incommensuration.
However only the doping dependence of the magnetic incommensuration has been carefully
studied (see \cite{tran_rev} and references therein).
The experiments show that the spin incommensuration is proportional to the doping $x$ for $x < 1/8$
and then saturates.
On the other hand 
charge peaks have been observed only for a narrow range
of dopings and only in stripe stabilized cuprates (in
La$_{1.5}$Nd$_{0.4}$Sr$_{0.1}$CuO$_4$ \cite{Nd}, La$_{1.475}$Nd$_{0.4}$Sr$_{0.125}$CuO$_4$ \cite{john},
La$_{1.45}$Nd$_{0.4}$Sr$_{0.15}$CuO$_4$ \cite{nie}, and
La$_{1.875}$Ba$_{0.125}$CuO$_4$ \cite{abba}).   While the observed charge incommensuration, $Q_{ch}$,
changes as a function of $x$ (and so is supportive of Scenario I), it does so weakly, i.e. the
change in $Q_{ch}$ is governed by $\delta Q_{ch} = c \delta x$ with $c\sim 0.5$.
One might then want to conclude that Scenario II remains a possibility, at least for dopings in
a narrow window about $x\sim 0.125$.
The situation is similarly ambiguous for 
$\rm YBa_2Cu_3O_{6+x}$.
Here distinct measurements of the phonon anomaly support alternatively
Scenario I \cite{mook,keimer1} and Scenario II \cite{pint}.

Scenario II is viable away from 1/8 doping in another fashion.  In Ref. \onlinecite{john1}, the behaviour
of the incommensuration as a function of $x$ is explained by invoking stripe spacing disorder.  Modifying this approach,
we can imagine stripe spacing disorder producing changes in the charge incommensuration while the magnetic incommensuration
arises from the doped ladder quasi-coherent mode (and so again is parallel to the stripe).  
This then yields a incommensuration pattern
\be
{\bf Q^{II'}}_{ch} = (0,\pm 4\pi x), ~~~ {\bf Q^{II'}}_{s} = (\pi \pm 2\pi x,\pi) ,
\ee
which gives Scenario I's charge incommensuration but Scenario II's magnetic incommensuration.

Independent of either scenario, our modeling efforts show the value of taking into account the doped
region of the copper-oxide planes.  In coupling together the ladders, we employed {\it antiferromagnetic}
couplings but were nonetheless able to explain the appearance of incommensurate order and the corresponding
$\pi$-phase shift in magnetic order.  If the doped regions were instead considered inert, a
{\it ferromagnetic} coupling would have to be assumed between adjacent doped 
ladders \cite{kruger,vojta,uhrig,carlson}.

While we have focused on magnetism here, models
of coupled ladders have promising superconducting properties.  It has already been established
that a model of a uniform array of coupled half-filled
ladders possesses narrow arcs of quasi-particles which have an instability towards 
d-wave superconductivity \cite{konricetsv}.  Ultimately this is a consequence of the presence of 
nascent d-wave superconducting order on the component ladders \cite{Fisher,Fabrizio}.
However these arcs are highly anisotropic with an alignment parallel to the ladders.
A question that then should be asked is whether a model of an array of ladders with
alternating doping
can do better.
Can such a model produce arcs aligned at $45^o$?  If it could, it would fill
in an important piece of the puzzle of how 1/8-doped LBCO can exhibit both stripe order
together with nodal quasi-particles \cite{ARPES}.

\acknowledgments
This work was
supported by the EPSRC under grant GR/R83712/01 (FHLE), the DOE under
contract DE-AC02-98 CH 10886 (AT and RMK) .
FHLE acknowledges the support from Theory Institute for Strongly
Correlated and Complex Systems at BNL and NSF DMR 0240238.
We are grateful to J. Tranquada for
discussions and interest in the work.

\appendix

\section{Susceptibilities of the Doped Ladders}

We extract  the susceptibilities of doped ladder  from a field theoretic
reduction of the ladders (Ref. \cite{FabRob}). The corresponding field theory
takes the form of the SO(6) Gross Neveu model supplemented by a U(1) Luttinger liquid 
describing the charge sector.  Though such a description captures low energy features of the 
system, it leaves us with ambiguities concerning the amplitudes of the correlation functions. 
It is these ambiguities which give rise to the possibility of different ordering scenarios 
described in the text. 

The field theory predicts the following general form for the susceptibilities:
\begin{eqnarray}\label{eAi}
\chi^{\rm d}_0(\omega,q) &=& \frac{3}{8} A_{11} J_1 (\omega, q);\cr\cr
\chi^{\rm d}_{2k_{F+}}(\omega,q) &=& \frac{3}{8} A_{12} \bigg(J_2 (\omega, q+2K_{F+}) + J_2 (\omega, q-2K_{F+})\bigg);\cr\cr
\chi^{\rm d}_{2k_{F-}}(\omega,q) &=& \frac{3}{8} A_{12} \bigg(J_2 (\omega, q+2K_{F-}) + J_2 (\omega, q-2K_{F-})\bigg);\cr\cr
\chi^{\rm d}_{k_{F+}-k_{F-}}(\omega,q) 
&=& \frac{3}{8} A_{31} \bigg(J_1 (\omega, q+K_{F-}-K_{F+}) + J_1 (\omega, q-K_{F-}+K_{F+})\bigg);\cr\cr
\chi^{\rm d}_{k_{F+}+k_{F-}}(\omega,q) 
&=& \frac{3}{8} A_{32} \bigg(J_3 (\omega, q+K_{F-}+K_{F+}) + J_3 (\omega, q-K_{F-}-K_{F+})\bigg).
\end{eqnarray}
In these expressions $A_{ij}$ are amplitudes with dimensionality of momentum that
are determined by short-distance physics.
On the other hand $J_i$'s are functions dependent on long-distance physics that arise from the form of the matrix elements
of the spin operators in the $SO(6)$ Gross-Neveu model. Different choice of the amplitudes, $A_{ij}$,
determine what ordering scenario is realized.

The imaginary part of $J_1$ takes the form 
\begin{eqnarray}\label{eAii}
&& {\rm Im } J_1(\omega,q) = \frac{8v_F\tilde{q}^2}{(\omega^2-\tilde{q}^2)^{3/2}}
\frac{\theta(\omega-\sqrt{\tilde{q}^2+4m^2})}{(\omega^2-\tilde{q}^2-4m^2)^{1/2}}\cr\cr
&&\ \times\ \exp\bigg[\int^\infty_0 \frac{dx}{x}
  \frac{G_c(x)}{s(x)}\big(1-c(x)\cos(\frac{\theta_{12}x}{\pi})\big)\bigg],
\end{eqnarray}
where $\theta_{12}$ is given by 
\begin{equation}\label{eAiii}
\theta_{12} = \cosh^{-1}(\frac{\omega^2-\tilde{q}^2-2m^2}{2m^2}),
\end{equation}
and $\tilde{q} = v_F q$ where $v_F$ is the Fermi velocity of electrons in either the bonding
or antibonding bands.  Here $s(x)/c(x) \equiv \sinh(x)/\cosh(x)$ and
$$
G_c(x) = \frac{e^{x/2}-1}{s(x)}.
$$
The imaginary parts of $J_2$ and $J_3$ can be expressed more compactly as integrals over
hypergeometric functions:
\begin{eqnarray}\label{eAv}
J_2(\omega,q) &=& \frac{v_F}{m^2}\int_{-\infty}^\infty d\theta 
\frac{s^2(\theta)}{\left[c(\theta)\right]^\frac{4-K}{2}}
F\big(1-\frac{K}{4},1-\frac{K}{4},1,
\frac{(\omega + \ri 0)^2 -\tilde{q}^2}{4m^2c^2(\theta)}\big)\cr
&& \hskip 1in
\exp\Big\{\int^\infty_0\frac{dx}{x}\frac{G_s(x)}{s(x)}[1-c(x)\cos(\frac{2\theta x}{\pi})]\Big\};\cr\cr
J_3(\omega,q)&=&
\frac{v_F}{2g^2m^2}\exp\Big(-2\int^\infty_0 \frac{dx}{x}
  \frac{G_v(x)}{s(x)}s^2(\frac{x}{4})\Big)
F\bigg(1-\frac{K}{4},1-\frac{K}{4},1,
\frac{(\omega + \ri 0)^2-\tilde{q}^2}{2m^2}\biggr)\cr\cr
&&\hskip -.5in +\frac{2^{K/4}v_F}{\pi m^2}\int_{-\infty}^\infty d\theta
\frac{1}{\left[c(\theta)\right]^\frac{4-K}{2}}\Big[\frac{s(2\theta)^2}{2c(2\theta)^2} +
\frac{s(\theta)^2}{c(2\theta)^2}\Big]
F\biggl(1-\frac{K}{4},1-\frac{K}{4},1,
\frac{(\omega + \ri 0)^2-\tilde{q}^2}{4m^2c^2(\theta)}\biggr)\cr\cr
&& \hskip .5in 
\times \exp\Big\{\int^\infty_0\frac{dx}{x}\frac{G_v(x)}{s(x)}[1-c(x)\cos(\frac{2\theta  x}{\pi})]\Big\},
\end{eqnarray}
where
$$
G_v(x) = \frac{2}{1-e^{-2x}}\bigg( e^{-2x}(1-e^{x/2})-e^{-5x/2}(1-e^{2x})\bigg).
$$
Here $K$ is the Luttinger parameter governing the gapless total charge mode of the doped ladder
and $g$ is a constant given by
$$
g^2 = \frac{2\sqrt{\pi}\Gamma(7/4)}{3\Gamma(5/4)}.
$$
To evaluate the real parts of $J_a$ ($a=1,2,3$) we Kramers-Kronig transform the above expressions for ${\rm Im}J$:
\begin{eqnarray}\label{eAvi}
{\rm Re} J(\omega,q) = \frac{1}{\pi}\int^{D}_{-D} d\omega' \frac{{\rm Im}J(\omega',q)}{\omega-\omega'}.
\end{eqnarray}
We equip the transformation with a cutoff, $D\sim v_F/a$, to reflect the fact the imaginary parts of $J_2$ and $J_3$
are only accurate representations of the low energy sector of the ladders.  To determine the real parts
of $J_{2/3}$ we perform the transformation over a frequency interval roughly corresponding
to this sector.  Our results, however, are insensitive to the exact value of $D$.

\subsection{Scenario I: Ordering Perpendicular to the Ladders}

To be able to produce the analysis of ordering perpendicular to the ladders (Scenario I -- Figs. 5, 7, and 8)
we had to fix the parameters $v_F A_{ij}/m^2$ and determine the value of spin gap $m$. In Ref. \cite{FabRob}, 
we did this 
through a comparison with an RPA analysis of a Hubbard ladder with an onsite U repulsion. 
The value of the spin gap $m$ is related to the bandwidth, $t$, the Hubbard interaction, $U$, and the Fermi velocity $v_F$. 
For $t \approx U$ we chose $t = 4ev$ and for $v_F = 350meVa$ (where a is the lattice spacing).  $v_F$ is not
readily available as its (strong) renormalization due to interactions is a two-loop effect.  But the value
we employed is commensurate with what is measured in the cuprates \cite{radu}.  As discussed in Ref. \cite{FabRob},
knowledge of $t$ alone is enough to fix the value of the gap, $m=26{\rm meV}$, using a field theory
analysis for doped ladders \cite{klls} together with the values of the gap on the ladder as determined from DMRG 
at half-filling \cite{eric,weihong}.  From this same field theory analysis, the Luttinger parameter
can be determined as $K=0.945$.  In Figs. 5 and 7 we couple the ladders together with
a strength just below that of the critical interladder coupling as determined from
our RPA analysis -- here $J_{crit} = 16.05$meV.
We choose the value of $J$ on the undoped ladders to be $100meV$ so as to match experimental observations
of the location of the neck of the hourglass describing the evolution of excitations in $La_{1.875}Ba_{0.125}CuO_4$ \cite{tran_nat}.
Furthermore to partially mimic the broadening seen in experiment, we broadened the spectral function of
the undoped ladders by assuming a lifetime of $0.1J$. 

The constants, $A_{ij}$, that appear in Eqn. (\ref{eAi}) are not determined by the field theory
treatment itself but must be accessed through separate considerations.  In Ref. \cite{FabRob} we,
through a comparison with a RPA analysis of a Hubbard ladder with an onsite U repulsion, were able
to provide tentative values for the $A_{ij}$'s.

\subsection{Scenario II: Ordering Parallel to the Ladders}

Since the amplitudes $A_{ij}$ are determined by processes with energies of the order of the bandwidth, 
we have a liberty of choice. For Scenario I we have chosen the high energy physics as in a simple doped 
Hubbard ladder with a point-like interaction and $U \sim t$. One can imagine that some other lattice realization 
generates a set of amplitudes such that the spectral weight associated with
the quasi-coherent spin excitation dominates.  To develop this scenario,
we thus focus on this excitation to the exclusion of contributions coming from
two excitation scattering continua.  Specifically we set $A_{11}=A_{12}=A_{31}=0$,
leaving only $A_{32}$ finite, and
take $J_3$ to equal
\begin{equation}\label{eAvii}
J_3(\omega,k) = 
\frac{v_F}{2g^2m^2}\exp\bigg(-2\int^\infty_0 \frac{dx}{x}
  \frac{G_v(x)}{\sinh(x)}\sinh^2(\frac{x}{4})\bigg)
F\bigg(1-\frac{K}{4},1-\frac{K}{4},1,
\frac{\omega^2-\tilde{k}^2}{2m^2}\biggr).
\end{equation}
As such, $J_3$ now represents a coherent mode broadened by the presence of gapless charge excitations.

For this scenario, in order to produce the spin response at constant energy in Fig. 6 and
the integrated intensity in Fig. 7, we must choose the ratio of $v_f/m$ to be sufficiently large
in order to guarantee Scenario II prevails over Scenario III.  To ensure this we chose
$v_f = 250 {\rm meV} a$, and the lattice bandwidth to be $t = 1000meV$ (and thus via Refs. \cite{klls,eric,weihong},
$m = 0.0065t = 6.5meV$).  Both the parameters used for the undoped ladder and
the Luttinger parameter describing charge excitations on the doped ladders where the same as in Scenario I (appendix A1).

In order to produce the results displayed in the left hand side of Fig. 8 
(discussing the strength of the inelastic signal in $\rm La_{1.86}Sr_{0.14}CuO_4$ at the 
incommensurate wave vector), we took instead $v_f = 360 {\rm  meV} a$ and $t=3000meV$.
This in turn moved the gap scale on the doped ladders to $m = 21 {\rm meV}$, high enough so that
it did not interference with the low energy signal marking nascent incommensurate order.


\begin{thebibliography}{99}
\bibitem{bedmull} J. G. Bednorz and K. A. M\"uller, Z. Phys. B: Condens. Matter {\bf 64}, 189 (1986).
\bibitem{moodenbaugh} A. R. Moodenbaugh, Y. Xu, M. Suenaga, T. J. Folkerts, 
and R. N. Shelton, Phys. Rev. B {\bf 38} (1988) 4596.
\bibitem{fujita} M. Fujita, H. Goka, and K. Yamada, J. M. Tranquada, 
and L. P. Regnault, Phys. Rev. B {\bf 70} (2004) 104517.
\bibitem{tran_nat} J. M. Tranquada, H. Woo, T. G. Perring, H. Goka, G. D. Gu, G. Xu, M. Fujita, and K. Yamada,
Nature {\bf 429}, 534 (2004).
\bibitem{abba} P. Abbamonte, A. Rusydi, S. Smadici, G. D. Gu, G. A. Sawatzky, and
D. L. Feng, Nature Phys. {\bf 1}, 155 (2005).
\bibitem{tran_rev} J. M. Tranquada, {\it Neutron Scattering Studies of Antiferromagnetic Correlations in Cuprates}
appearing in {\bf Handbook of High-Temperature Superconductivity}, ed. J. R. Schrieffer and J. S. Brooks, Springer New York 
(2005).
\bibitem{nickelates}Stripe phases are also found in allied compounds to the cuprates, for example
the nickelates, $La_{2-x}Sr_xNiO_{4+\delta}$.  See for example J. M. Tranquada, D. J. Buttrey, V. Sachan, and
J. E. Lorenzo, Phys. Rev. Lett. {\bf 73}, 1003 (1994).
\bibitem{dai} P. Dai, H. A. Mook, R. D. Hunt, and F. Do\v{g}an, Phys. Rev. B {\bf 63}, 054525 (2001). 
\bibitem{mook} H. A. Mook, P. Dai, F. Do\v{g}an, and R. D. Hunt, Nature {\bf 404}, 729 (2000).  
\bibitem{keimer1} V. Hinkov, S. Pailh\`es, P. Bourges, Y. Sidis, A. Ivanov, A. Kulakov, C. T. Lin, D. P. Chen, 
C. Bernhard, and B. Keimer Nature {\bf 430}, 650 (2004).
\bibitem{hinkov} V. Hinkov, P. Bourges, S. Pailhès, Y. Sidis, A. Ivanov, C. D. Frost, 
T. G. Perring, C. T. Lin, D. P. Chen, and B. Keimer, Nature Physics 3, 780 (2007).
\bibitem{mook1} See Ref.(\onlinecite{mook}) above for evidence to this effect.  We do note however
that Ref. \cite{pint} has reported observations of this same phonon anomaly indicating that
the incommensurate spin modulation is found along the stripes.
\bibitem{pint} L. Pintschovius, W. Reichardt, M. Kl\"aser, T. Wolf, and H. v. L\"ohneysen, Phys. Rev. 
Lett. {\bf 89}, 037001 (2002).
\bibitem{shirane} K. Yamada, C. H. Lee, K. Kurahashi, J. Wada, S. Wakimoto, S. Ueki, H. Kimura, Y. Endoh,
S. Hosoya, G. Shirane, R. J. Birgeneau, M. Greven, M. A. Kastner, and Y. J. Kim, Phys. Rev. B {\bf 57}, 6165 (1998).
\bibitem{shirane1} M. Fujita, K. Yamada, H. Hiraka, P. M. Gehring, S. H. Lee, S. Wakimoto, and G. Shirane,
Phys. Rev. B {\bf 65}, 064505 (2002).
\bibitem{kruger} F. Kr\"uger and S. Scheidl, Phys. Rev. B {\bf 67}, 134512 (2003).
\bibitem{vojta} M. Vojta and T. Ulbricht, Phys. Rev. Lett. {\bf 93},
  127002 (2004).
\bibitem{uhrig}
 G.S. Uhrig, K.P. Schmidt and M. Gr\"uninger,  Phys. Rev. Lett. {\bf
 93}, 267003 (2004). 
\bibitem{carlson} D. X. Yao, E. W. Carlson, D. K. Campbell, Phys. Rev. B 73, 224525 (2006);
D. X. Yao, E. W. Carlson, Phys. Rev. B 77, 024503 (2008).
\bibitem{note} In particular, the low energy field theoretic treatment that we employ to characterize
the spin response of the doped ladder does not uniquely fix the real part of the susceptibility, a quantity
dependent upon high energy non-universal physics.
\bibitem{xu} Guangyong Xu, J. M. Tranquada, T. G. Perring, G. D. Gu, M. Fujita, and K. Yamada,
Phys. Rev. B 76, 014508 (2007).
\bibitem{kiv}
M. Granath, V. Oganesyan, S.A. Kivelson, E. Fradkin and V. Emery,
Phys. Rev. Lett. {\bf 87}, 167011 (2001);
M. Granath, V. Oganesyan, D. Orgad and S.A. Kivelson,
Phys. Rev. B{\bf 65}, 184501 (2002).
\bibitem{ARPES}
T. Valla, A.V. Fedorov, J. Lee, J.C. Davis and G.D. Gu, Science {\bf  314}, 1914 (2006);
M.R. Norman, A. Kanigiel, M. Randeria, U. Chatterjee and
J.C. Campuzano, Phys. Rev. B 76, 174501 (2007).
\bibitem{Fabrizio}
M. Fabrizio, Phys. Rev. B {\bf 48}, 15838 (1993).
\bibitem{Fisher} 
H. L. Lin, L. Balents and M. P. A. Fisher, Phys. Rev. B {\bf 58}, 1794
(1998). 
\bibitem{AFK} 
E. Arrigoni, E. Fradkin and S. Kivelson, Phys. Rev. B{\bf 69}, 214519
(2004). 
\bibitem{konricetsv}
R. M. Konik, T. M. Rice, and A. M. Tsvelik, Phys. Rev. Lett. 96, 086407 (2006).
\bibitem{lake}
S. Notbohm, P. Ribeiro, B. Lake, D. A. Tennant, K. P. Schmidt, G. S. Uhrig, 
C. Hess, R. Klingeler, G. Behr, B. Büchner, M. Reehuis, R. I. Bewley, C. D. Frost, P. Manuel, and R. S. Eccleston, 
Phys. Rev. Lett. {\bf 98}, 027403 (2007).
\bibitem{lake1} B. Lake, G. Aeppli, T. E. Mason, A. Schröder, D. F. McMorrow, K. Lefmann, M. Isshiki, 
M. Nohara, H. Takagi, and S. M. Hayden, Nature {\bf 400} 43 (1999).
\bibitem{christensen} N. B. Christensen, D. F. McMorrow, H. M R\o nnow, B. Lake, S. M. Hayden, G. Aeppli, 
and T. G. Perring, Phys. Rev. Lett. {\bf 93}, 147002 (2004).
\bibitem{vignolle} B. Vignolle, S. M. Hayden, D. F. McMorrow, H. M. R\o nnow, B. Lake, C. D. Frost, and
T. G. Perring, Nature Physics {\bf  3}, 163 (2007).
\bibitem{tran_lsco} J. M. Tranquada, C. H. Lee, K. Yamada, Y. S. Lee, L. P. Regnault, and H. M. R\o nnow,
Phys. Rev. B {\bf 69}, 174507 (2004).
\bibitem{barnes} T. Barnes and J. Riera, Phys. Rev. B {\bf 50}, 6817 (1994).
\bibitem{SU05}
K.P. Schmidt and G.S. Uhrig, Mod. Phys. Lett. {\bf B19}, 1179 (2005).
\bibitem{ZEK} O. Zachar, S. A. Kivelson and V. J. Emery, Phys. Rev. B{\bf 57}, 1422 (1998). 
\bibitem{Nd} T. Niem\"oller, N. Ichikawa, T. Frello, H. H\"nnerfeld, N. H. Andersen, S. Uchida, J. R. Schneider, 
J. M. Tranquada, Euro. Phys. J B {\bf 12}, 509 (1999).
\bibitem{john} J. M. Tranquada, B. J. Sternlieb, J. D. Axe, Y. Nakamura, and S. Uchida, Nature {\bf 375}, 561 (1995).
Phys. Rev. Lett. {\bf 85}, 1738 (2000).
\bibitem{nie} N. Ichikawa, S. Uchida, J. M. Tranquada, T. Niem\"oller, P. M. Gehring, S.-H. Lee, and J. R. Schneider, 
Phys. Rev. Lett. {\bf 85}, 1738 (2000).
\bibitem{john1} J. M. Tranquada, N. Ichikawa, and S. Uchida, Phys. Rev. B {\bf 59}, 14712 (1999).
\bibitem{FabRob} F. H. L. Essler and R. M. Konik, Phys. Rev. B {\bf 75},
  144403 (2007). 
\bibitem{klls} R. M. Konik, F. Lesage, A. Ludwig, and H. Saleur, Phys. Rev. B {\bf 61}, 4983 (2000).
\bibitem{radu} R. Coldea, S. M. Hayden, G. Aeppli, T. G. Perring, C. D. Frost, T. E. Mason, S.-W. Cheong, and Z.
Fisk, Phys. Rev. Lett. {\bf 86}, 5377 (2001).
\bibitem{eric}
E. Jeckelmann, D.J. Scalapino and S.R. White, Phys. Rev. B{\bf 58},
9492 (1998). 
\bibitem{weihong}
Z. Weihong, J. Oitmaa, C.J. Hamer and R.J. Bursill,
J. Phys. Cond. Matt. {\bf 13}, 433 (2001).

\end{thebibliography}
\end{document}